\documentclass[12pt]{article}

\usepackage{amsmath,amssymb,wasysym}
\usepackage{graphicx,color}

\def\cN{{\mathcal{N}}}

\def\cM{{\mathcal{M}}}
\def\cT{{\mathcal{T}}}

\def\bz{\bar{z}}

\def\be{\beta}
\def\al{\alpha}
\def\ga{\gamma}

\def\r2{{\sqrt{2}}}
\def\h{{\eta}}

\def\h0{\hat{h}}
\def\Vr0{\hat{V}_{r}}
\def\Vp0{\hat{V}_{\phi}}

\def\r2{\sqrt{2}}
\def\be{\begin{equation}}
\def\ee{\end{equation}}

\def\sepy{|\bf{y}-\bar{\bf{y}}|}
\def\bA{{\bf A}}
\def\bB{{\bf B}}

\newcommand{\Letter}{
\setlength{\textwidth}{16.5cm}
   \setlength{\textheight}{22.6cm}
    \hoffset=-0.6in
\voffset=-2.1cm }

\Letter

\begin{document}
\begin{titlepage}
\begin{flushright}
DAMTP-08-105\\
MAD-TH-08-15\\
SU-ITP-09-02
\end{flushright}
\begin{centering}
%^{\rm full}
\vspace{.3in}

{\Large {\bf Inflation on an Open Racetrack}}

%\bigskip

\vspace{.3in}

Heng-Yu Chen${}^1$, Ling-Yan Hung${}^2$, and Gary Shiu${}^{1,3}$ \\
\vspace{.2 in}
${}^{1}$\textit{Department of Physics, University of Wisconsin,
Madison, WI 53706, USA}
%
%\vspace{.2in}
%\vskip 3pt
\vskip 3pt
%\vspace{.2in}
\vspace{.1 in}
${}^{2}$\textit{DAMTP, Centre for Mathematical Sciences,
University of Cambridge, \\
Wilberforce Road,
Cambridge CB3 0WA, UK}
%\vspace{.2in}
%and \\
%\vspace{.2in}
%\vspace{.1 in}
\vskip 3pt
${}^3$\textit{Department of Physics and SLAC, Stanford University,
Stanford, CA 94305, USA}

\bigskip \bigskip

{\bf Abstract}

\vspace{.1in}

We present a variant of warped D-brane inflation by
incorporating
multiple
 sets of
holomorphically~\!-\!~embedded
D7-branes
involved in moduli stabilization with extent into a warped throat.
The
resultant
D3-brane motion
depends on the D7-brane configuration and
the relative position of the D3-brane in these backgrounds.
The
non-perturbative moduli stabilization superpotential
takes the racetrack form, but the additional
D3-brane open string moduli dependence
provides more flexibilities in model building.
For concreteness, we
consider
D3-brane motion in the warped deformed conifold with the presence of multiple D7-branes, and  derive
 the scalar potential valid for the entire throat.
By explicit tuning of the microphysical parameters, we
obtain inflationary trajectories near an inflection point for various
D7-brane
 configurations.
Moreover, the open racetrack potential admits approximate Minkowski vacua before uplifting.
We demonstrate with a concrete D-brane inflation model where the Hubble scale
during inflation
can  exceed the gravitino mass.
Finally, the multiple sets of D7-branes present in this open racetrack setup also provides a mechanism to stabilize the D3-brane to metastable vacua in the intermediate region of the warped throat.
\end{centering}

\vfill

\begin{flushleft}
\today
\end{flushleft}

\end{titlepage}

\section{Introduction and Summary}
\paragraph{}
The inflationary paradigm \cite{inflation} offers a compelling solution to some of the most puzzling features of standard big-bang cosmology, notably the flatness and the horizon problems.
Since its inception, a myriad of effective field theory based models of inflation have been proposed.
In the coming decade, increasingly precise cosmological data will help to constrain the
``theory
 space" of inflation
to the extent that hard data may enable us to disfavor or even rule out
some of the popularly proposed models.
Thus, it is of interest to examine what are the generic predictions of inflation, and what merely are the
consequences of specific models.

In this regard, a particularly important feature of inflation is its ultra-violet sensitivity. This feature
is most clearly
exemplified by the fact that mass dimension six, Planck scale suppressed corrections to the inflation potential  can give order one contributions to the slow-roll parameters and hence significantly alter
the dynamics of inflation. Therefore, a truly predictive model of inflation would
require a consistent microscopic theory of quantum gravity, such as string theory, for completion.
It is then natural to ask if  embedding inflation into a microscopic framework
can impose restrictions on the inflationary ``theory space",  thus enabling us to sharpen the predictions of inflation.
This question is especially interesting in the context of string cosmology
since the construction and predictions
of string inflationary models are intimately tied to the microphysics of moduli stabilization.
This
sensitivity on the details of string compactifications  manifests in various forms in recent studies of
 string inflation, e.g.,
in addressing the supergravity $\eta$ problem \cite{KKLMMT,McGill1,Baumann:2007np,Krause:2007jk,Baumann1,HoloBulkeffects},
in limiting and extending the physical field range of the inflaton field \cite{Baumann:2006cd,Kobayashi:2007hm,Becker:2007ui,Kallosh:2007cc,Grimm:2007hs,Silverstein:2008sg,McAllister:2008hb,Cicoli:2008gp}, and in determining the end of inflation and multi-field effects \cite{Kecskemeti:2006cg,Shiu:2006kj,DeWolfe:2007hd,Easson:2007dh,Huang:2007hh,CGS1}. Given that string theory is not only a quantum theory of gravity, one might wonder if the  microphysics of moduli stabilization relevant for the
above
cosmological issues may also leave its footprints on particle physics.
If so, embedding inflation into string theory may
provide us with  some interesting and unexpected links between cosmology and particle physics data.

An interesting relation which illustrates this latter point was pointed out by \cite{KLbound}.
It was found that in the simplest inflationary models
based on the KKLT mechanism \cite{KKLT} of moduli stabilization, the Hubble scale
during inflation
is bounded by the present value of the gravitino mass, i.e., $H \apprle m_{3/2}$.
This relation, which
 ties the amplitude of primordial gravitational waves to the scale of supersymmetry breaking,
 appears to be rather generic\footnote{Its specific form may differ somewhat among known moduli stabilization mechanisms, see, e.g., \cite{CKLQ}.} among
the concrete moduli stabilization mechanisms studied to date.
Indeed, it has proven to be challenging to construct a natural string inflationary model with low scale supersymmetry and detectable primordial tensors.
As also discussed in the original work \cite{KLbound},
this gravitino mass bound can be evaded with  fine-tuning
and additional ingredients beyond the minimal scenario of
KKLT (for recent work elaborating further on this point, see \cite{CKLQ,KL2,AccidentalInflation,BO1,BO2,Abe:2008xu}).
These previous studies, however, focussed on models where the inflaton is a K\"ahler modulus.
As some of the most explicit string inflation models
often involve branes,
it is  worthwhile to revisit this issue in the context of brane inflation \cite{Dvali:1998pa} where the inflaton comes from the open string sector.

In this paper, we present a variant of warped D-brane inflation,
with the above motivation behind.
In the original KKLT scenario, a single set of holomorphically-embedded D7-branes
as minimally needed to stabilize the universal K\"ahler modulus was introduced.
However, there can a priori be non-perturbative contributions from more than one gauge sector.
Having multiple hidden gauge sectors is not atypical in string compactifications and in fact crucial in the ``racetrack" mechanism of moduli stabilization \cite{RacetrackUnpublished,Krasnikov:1987jj,Casas:1990qi,Taylor:1990wr}.
Thus, we consider a simple generalization of \cite{KKLMMT} by introducing multiple stacks of moduli-stabilizing D7-branes with extent into the warped throat where the inflationary D3-brane resides.
The resultant non-perturbative superpotential resembles the one appearing in racetrack inflation \cite{RaceTrack1}, but now with additional dependence on the
D3-brane moduli which we identify as the inflaton.
There are several advantages of considering this ``open racetrack" scenario.
As suggested in \cite{KLbound}, extending \cite{KKLT} to a racetrack form allows for  an approximate Minkowski vacuum (instead of an AdS minimum as in KKLT) before uplifting. The gravitino mass is then disentangled from
the height of the uplifted potential and thus the bound $H \apprle m_{3/2}$ can be evaded.
In comparison to inflation on a ``closed racetrack", brane inflation on an ``open racetrack" offers more flexibilities in model building.
In particular, while a racetrack superpotential can circumvent the bound in \cite{KLbound}, explicit inflationary model building realizing the solution in \cite{KLbound}
with closed string moduli has proven to be non-trivial \cite{CKLQ,KL2,AccidentalInflation,BO1,BO2,Abe:2008xu}
since the inflaton field is in the same K\"ahler sector  being stabilized by the racetrack potential.   In some constructions,
additional terms (or novel moduli dependence\footnote{For example, in \cite{BO2,Abe:2008xu}, the gauge kinetic function of the worldvolume gauge fields on some exotic D-branes found in \cite{MarchesanoShiu} was used.}) in the non-perturbative superpotential
and/or extra global symmetries were invoked to ensure that closed string racetrack
inflation can take place.
As we shall see, having the inflaton in the open string sector separates the tuning required for
inflation  from that of the minimum of the pre-uplifted moduli stabilization potential.
This flexibility allows us to consider several scenarios depending on the
configuration of D7-branes and the relative position of the mobile D3-brane in these backgrounds.
In some cases, the forces on the D3-brane exerted by different stacks of
D7-branes balance off each other, resulting in a metastable D3-brane vacuum at a finite tunable distance from the infrared end of the warped throat. Inflation can then be viewed as accidental in this setup when additional uplifting sources and appropriate choices of compactification data are chosen such that the metastable minimum turns into an inflection point.
Finally, while the main focus of this work is inflation, our mechanism of obtaining metastable D3-brane vacua may have more general applicability, e.g., in particle physics issues when the Standard Model particles and/or hidden sector fields are localized on the worldvolume of D-branes.
We leave the studies of these applications to future work.

This paper is structured as follows. In Section \ref{NPsuperpotential}, we discuss in detail the
non-perturbative potential generated by gaugino condensates on multiple stacks of D7-branes
in a warped throat. In Section \ref{AnExample}, we use the D7 brane embedding of \cite{Kuperstein} to illustrate how a single field inflatonary potential, valid for the entire deformed conifold, can
 arise from such racetrack superpotential.
In Section \ref{KLbounddiscussion}, we review the bound pointed out in \cite{KLbound} and explain how it can be evaded in the open racetrack setup. In Section \ref{Results}, we present  several scenarios of D7-brane configurations and some explicit parameter sets for each case
such that a sustained period of inflation can take place,
including an example where the bound \cite{KLbound} can be circumvented. We relegate some technical details to a series of appendices.

\section{Non-Perturbative  Potential in Warped Throats}\label{NPsuperpotential}

\paragraph{}
In this section we shall begin by
collecting some results of
flux compactifications \cite{fluxreviews,verlinde99,Dasgupta:1999ss,Greene:2000gh,Becker:2000rz,GKP}, in particular, the effective 4D $\cN=1$ supergravity  action that are useful for describing
warped brane inflation. This review also serves to set  our notation. We
will also present a detailed form of the
scalar potential
valid in the entire
warped deformed conifold.

The $\mathcal{N}=1$ F-term scalar potential in supergravity is given by
\begin{eqnarray}
V_{F}(\rho,\bar{\rho}, z^{\alpha},\bar{z}^{\alpha}) &=& e^{\kappa^2 \mathcal{K}} \big[\mathcal{K}^{\Sigma\Omega}D_{\Sigma}W\overline{D_{\Sigma}W}- 3\kappa^2|W|^2\big]\,,~~~\kappa^2=M_P^{-2} \equiv 8\pi G\,,\label{DefFscalarpot1}  \\
D_{\Sigma}W &=& \partial_{\Sigma}W + \kappa^2 (\partial_{\Sigma}\mathcal{K})W\,.\label{DefCovDeriv}
\end{eqnarray}
Here the indices $\{Z^{\Sigma}\}\equiv \{\rho, z^\alpha\}$
where $z^\alpha$ are the complex coordinates of the compact space, and $\rho= \sigma + i\xi$ is the complex K\"ahler modulus, whereas $W\equiv W(z^\alpha,\rho)$ is the holomorphic superpotential.
For simplicity, we shall consider in this paper the situation where the compact Calabi-Yau space only has a single K\"ahler modulus.
The K\"ahler potential  $\mathcal{K}$ in the presence of the D3 brane is given by \cite{GKP,DeWolfeGiddings}\footnote{The Kahler potential is modified in the strongly warped limit.
 We refer the readers to some recent work \cite{Giddings:2005ff, Frey:2006wv, Burgess:2006mn, Shiu:2008ry,Douglas:2008jx,Frey:2008xw,Marchesano:2008rg} for a discussion of the subtle issues involved in the derivation of warped Kahler potential.}:
\begin{eqnarray}
\kappa^2\mathcal{K}(z^\alpha,\bz^\alpha,\rho,\bar{\rho}) &=& -3\log \left[ \rho + \bar{\rho} - \gamma k(z^{\alpha}, \bar{z}^{\alpha}) \right]  \equiv -3\log U(z^\alpha,\bz^{{\alpha}},\rho,\bar{\rho}), \label{Deftotalkahlerpot}
\end{eqnarray}
where the constant $\gamma = \frac{\sigma_0 T_3}{3M_P^{2}}$ with $\sigma_0$ the stabilized value of $\sigma$ when the D3 brane assumes its stabilized position, and $k(z^\alpha,\bar{z}^{\alpha})$ is the K\"ahler potential of the ambient Calabi-Yau space where the mobile D3 brane moves.
This allows us to further expand the expression for the $F$-term scalar potential (\ref{DefFscalarpot1}) in terms of the inverse K\"ahler metric of the ambient Calabi-Yau, and the derivatives with respect to the local coordinates and the K\"ahler modulus \cite{McGill1}:
{\small \begin{eqnarray}
V_F(z^\al,\bz^\al,\rho,\bar{\rho}) &= & \frac{\kappa^2}{3[U(z^\alpha,\bz^{{\alpha}},\rho,\bar{\rho})]^2} \left\{ \left[ U(z^\alpha,\bz^{{\alpha}},\rho,\bar{\rho}) +
\gamma k^{\gamma\bar\delta} k_{\ga} k_{\bar{\delta}} \right] |W_{,\rho}|^2 - 3
\left( \overline{W}W_{,\rho} + W\overline{W}_{,\bar{\rho}} \right) \right\}
\nonumber\\
& +& \frac{\kappa^2}{3[U(z^\alpha,\bz^{{\alpha}},\rho,\bar{\rho})]^2} \left\{ \left(
k^{\al\bar\delta}k_{\bar\delta}\overline{W}_{,\bar{\rho}}W_{,\al} +
k^{\bar{\al}\delta}k_{\delta}W_{,\rho}\overline{W}_{,\bar{\al}} \right) +
\frac{1}{\gamma} k^{\al\bar\beta}W_{,\al}\overline{W}_{,\bar{\beta}} \right\} \,.
\label{explicitVF}
\end{eqnarray}}
$\!\!$Notice that the second line in (\ref{explicitVF}) comes strictly from the dependence of the superpotential $W(z^\alpha,\rho)$ on the mobile D3 position, which in turns generates a non-trivial potential for its trajectory.

The superpotential we shall consider in this paper takes the form
\begin{equation}
W(z^\alpha,\rho) = W_0 + A(z^\alpha)e^{-a_1\rho} + B(z^\alpha)e^{-a_2\rho}\label{DefW2stack},
\end{equation}
where $W_0 \equiv \int G\wedge \Omega$ is the Gukov-Vafa-Witten flux superpotential \cite{GKW}. $W_0$ depends on the dilaton-axion and the complex structure moduli, which we assumed to be fixed by the bulk fluxes \cite{GKP}. 
We shall further assume that in the following we can rotate the phase of the flux superpotential such that $W_0\in {\mathbb{R}}$. 
The remaining contributions in $W(z^\alpha,\rho)$ are non-perturbative contributions arising from two separate Euclidean D3 instantons or D7 gaugino condensates, which we shall discuss in detail momentarily. The superpotential (\ref{DefW2stack}) resembles the one used in ``Racetrack Inflation'' \cite{RaceTrack1}. However, in contrast with such model where the complex K\"ahler modulus $\rho$ acts as the inflaton, the canonical inflaton here is  identified with the radial position of a mobile D3 brane.
The functions $A(z^\alpha)$ and $B(z^\alpha)$ typically depend holomorphically on the complex structure moduli (which we assume to be stabilized perturbatively by the flux potential $W_0$) and also on the position moduli $\{z^{\alpha}\}$ of the mobile D3 brane.

In the case of a singular conifold, the explicit dependences on the D3 position for $A(z^\alpha)$ and $B(z^\alpha)$ has been calculated in \cite{Gaugino1} (see also earlier work \cite{Ganor,Berg}), and are given by
\begin{equation}\label{DefAzBz}
A(z^\al) = A_0 \left[ \frac{f_1(z^\al)}{f_1(0)} \right]^{1/n_1} \,,~~B(z^\al) = B_0 \left[ \frac{f_2(z^\al)}{f_2(0)} \right]^{1/n_2}\,,~~a_{1,2}=\frac{2\pi}{n_{1,2}}\,.
\end{equation}
Here $n_{1,2}>1$ is the number of D7s (or $n_{1,2}=1$ for Euclidean D3 instanton) on each brane stack.
The precise values of the complex constants $A_0$ and $B_0$ depend on the stabilized complex structure moduli as well as the dynamical scales on the D7-branes; whereas the holomorphic functions $f_1(z^{\alpha})$ and $f_2(z^{\alpha})$ are the embedding functions of the supersymmetric four cycles wrapped by the D7-branes or the
Euclidean D3-branes.
The dependence on $z^{\alpha}$ essentially comes from the fact that the conifold throat is attached to a compact bulk manifold, the mobile D3 brane backreacts on the holomorphic four cycle wrapped by the moduli-stabilizing D7s. This in turns induces a small but non-trivial force on D3.
An inflationary phase can be then generated if this force
balances off the contributions to the inflaton mass due to moduli stabilization effects
\cite{Baumann1}.

As in \cite{KKLT}, we have implicitly assumed that the translational modes of D7 branes have been stabilized (e.g., by fluxes) along the symmetry enhanced locus, which allows for gaugino condensation to take place.
In this paper, we shall consider a simplifed configuration such that the two stacks of D7s extending radially in the conifold do not intersect with each other. Thus, possible additional unsaturated fermionic zero modes that arise at the $7$-$7$ intersections can be avoided as they would cause the gaugino condensates, which are vital in stabilizing the K\"ahler modulus $\rho$, to vanish \cite{Ganor}.

Moreover, the functional dependence of the D7 gaugino condensate on the mobile D3 position in the singular conifold in fact  holds for the full deformed conifold, despite the fact that almost all of the known supersymmetric D7 embeddings in the singular conifold require extra worldvolume flux along the compact directions to remain supersymmetric in the full deformed conifold \cite{COSD7}. In Appendix \ref{GauginoCondensate}, we shall discuss from both the closed and the open string perspectives that the presence of such extra worldvolume gauge field does not affect the general form of D7 gaugino condensate given in (\ref{DefAzBz}).

There are several motivations for introducing a second stack of D7 gaugino condensate.
First,
a supersymmetric local minimum for the mobile D3 can be
obtained between the two stacks of moduli-stabilizing D7 branes where their forces on the D3 are balanced.
In contrast to the single stack case where the D3-brane is stabilized at the end of the throat \cite{DeWolfe:2007hd}, one can now dynamically stabilize the D3 position moduli in the intermediate region of the deformed conifold.
This suggests a variant of the warped
D-brane inflation scenario
\cite{stringcosmologyreviews}
with several qualitatively different properties.
In comparison with the usual inflationary models constructed from the simplest KKLT scenario, where a supersymmetric AdS minimum is uplifted to a dS one; having an extra gaugino condensate allows us to start with
a Minkowski vacuum instead. As discussed in \cite{KLbound}, this has an important consequence in
evading the constraint given in \cite{KLbound}, making it possible to obtain, at least in principle,
high scale inflation and low energy supersymmetry breaking simultaneously.
We shall discuss in more detail this gravitino mass bound in Section \ref{KLbounddiscussion}.
Furthermore,  the D3-vacua found here and in \cite{DeWolfe:2007hd},
together with the D7-brane vacua in flux compactifications explored in \cite{Gomis:2005wc}
 constitute a rich picture of an open string landscape.
The stabilization of open (versus closed) string moduli is of phenomenological
importance as the Standard Model is realized on the worldvolume of D-branes (and their intersections)
\cite{Blumenhagen:2005mu} in Type IIB flux compactifications \cite{fluxreviews}.
In particular, the masses of the light open string modes and their couplings to the closed string degrees of freedom depend on the distance between the branes and the local geometry at which the branes are stabilized. While the focus of the current work is inflation and so we will not dwell further on these issues,
we expect our results to have more general applicability including such questions of interest to
particle physics as well.

\section{An Explicit Example}\label{AnExample}
\paragraph{}
In this section we shall apply the general formulae considered above to the entire deformed conifold, and construct explicitly the single field inflaton potential from the racetrack superpotential (\ref{DefW2stack}), following some of the steps outlined in \cite{Baumann1}.
%%%%%%%%%%%%%%%%%%%%%%%%%%%%%%%%%%%%%%%%%%%%%%%%%%%%%%%%%%%%%%%%%%%%%%%%%%%%%%%
\subsection{Racetrack from Multiple Brane Stacks}
\paragraph{}
We shall in particular focus on the supersymmetric D7 brane embedding established in \cite{Kuperstein}, it is the only currently known embedding in the deformed conifold such that no additional world volume gauge flux is required \cite{COSD7}.
In order for the two stacks of D7 branes to be non-intersecting, we shall take the four cycles
on which they wrap to be given by
\begin{equation}\label{Deff1f2}
f_1(z^\alpha)=z_1-\mu_1\,,~~~f_2(z^\alpha)=z_1-\mu_2\,,~~\mu_1\,,\mu_2\in {\mathbb{R}}
\end{equation}
Thus, we have two stacks of D7s wrapping holomorphic four cycles of identical topology, but extend to different depths $|\mu_1|^{2/3}$ and $|\mu_2|^{2/3}$ into the warped throat\footnote{ Differing from \cite{Baumann1}, we have not restricted $\mu_1$ and $\mu_2$ to be positive, in fact the requirement of angular stability would give constraints on the value of $\mu_1, \mu_2$ and $A_0,B_0$.}.
Such embedding (\ref{Deff1f2}) preserves the $SO(3)$ subgroup of the $SO(4)$ isometry group for both the singular and the deformed conifold.

We can now apply the explicit expression (\ref{explicitVF}) to the superpotential generated by two stacks of D7 branes wrapping on the four cycles described above. First we note that the inverse metric for the full deformed conifold as derived in Appendix \ref{DefconifoldDetails} is given by:
{\small \begin{eqnarray}
&&k^{\bar{i}j}=\frac{r^3}{\frac{\partial^2 k(\tau)}{\partial\tau^2}}\left[\left(\delta^{\bar{i}{j}}-\frac{z_i\bz_j}{r^3}\right)+\frac{ r^3 \bB(\tau)}{\bA(\tau)}\left(\left(1-\frac{\epsilon^4}{r^6}\right)\delta^{\bar{i}{j}}
+\frac{\epsilon^2}{r^3}\left(\frac{z_i z_j+\bz_i\bz_j}{r^3}\right)-\left(\frac{z_i \bz_j+\bz_i z_j}{r^3}\right)\right)\right]\,,\nonumber\\
\label{DefConmetric2}\\
&&r^3\frac{\bB(\tau)}{\bA(\tau)}=\coth\tau\left(\frac{\partial^2_\tau k(\tau)}{\partial_\tau k(\tau)}-\coth\tau\right)\,,~~~i\,,\bar{j}=1,2,3\,.
\end{eqnarray}}
$\!\!\!$Here we have introduced the dimensionless radial coordinate $\tau$ which is valid for the entire deformed conifold, and is related to the usual radial coordinate $r$ via the relation $r^{3}=\epsilon^2\cosh\tau$ \cite{ConifoldNotes}. For more details on deformed conifold, see Appendix \ref{DefconifoldDetails}.
After some tedious but straightforward computations, one can obtain the F-term scalar potential for the embeddings defined in (\ref{Deff1f2}):
\begin{eqnarray}
V_{\textrm{F}} &=& V_{\textrm{KKLT}} + \Delta V, \label{DefVF2} \\
V_{\textrm{KKLT}} &=&  \frac{ \kappa^2  |S_{0}(z_1,\rho)|^2 }{3U(\tau,\sigma)^2}\left(U(\tau,\sigma) + (\epsilon^{4/3}\gamma) P(\tau) + 6{\mathrm {Re}}\left[\frac{W(z_1,\rho)}{S_{0}(z_1,\rho)}\right] \right)\,, \label{DefVKKLT} \\
\Delta V&=& \frac{\kappa^2|S_1(z_1,\rho)|^2}{3U(\tau, \sigma)^2}  \left(
\frac{T(z_1,\bar{z}_1)}{\gamma} + 2\coth\tau L(\tau) \textrm{Re}\left[\frac{S_0(z_1,\rho)}{S_1(z_1,\rho)} \left(\bar{z_1}- \frac{\epsilon^2}{r^3}z_1\right)\right]\right)\,.\label{DefDeltaV}
\end{eqnarray}
where $W(z_1,\rho)$ was defined in (\ref{DefW2stack}) with the embedding functions (\ref{Deff1f2}) substituted and the various other functions are given by:
{\small \begin{eqnarray}
U(\tau,\sigma)&=&2\sigma-\gamma k(\tau)\,,  \label{DefU} \\
S_0 (z_1,\rho) &=& a_1 A_0\left(1-\frac{z_1}{\mu_1}\right)^{1/n_1}e^{-a_1\rho} + a_2 B_0\left(1-\frac{z_1}{\mu_2}\right)^{1/n_2}e^{-a_2\rho}=-\frac{\partial W(z_1,\rho)}{\partial\rho}\,,\label{DefSm} \\
S_1 (z_1,\rho) &=& \frac{A_0}{n_1\mu_1}\left(1-\frac{z_1}{\mu_1}\right)^{1/n_1-1}e^{-a_1\rho} + \frac{B_0}{n_2\mu_2}\left(1-\frac{z_1}{\mu_2}\right)^{1/n_2-1}e^{-a_2\rho}=-\frac{\partial W(z_1,\rho)}{\partial z_1}\,,\label{DefSm1} \\
P(\tau)&=& \frac{3}{2}\frac{(\sinh(2\tau)-2\tau)^{4/3}}{2^{4/3}\sinh^2\tau}\,,~~~L(\tau)=\frac{3}{4}\frac{(\sinh(2\tau)-2\tau)}{\sinh^2\tau} \,,\label{DefKLtau}
\\
T(z_1,\bar{z_1}) &=& \frac{1}{\partial_\tau^2 k}\left\{(r^3-|z_1|^2) +  r^3\frac{\bB(\tau)}{\bA(\tau)} \left[(r^3-|z_1|^2)\left(1-\frac{\epsilon^4}{r^6}\right)
-\left(z_1-\frac{\epsilon^2}{r^3}\bar{z}_1\right)\left(\bar{z}_1-\frac{\epsilon^2}{r^3}z_1\right)\right] \right\}\,.\nonumber \label{DefTtau}\\
\end{eqnarray}}
$\!\!$Here we have seperated the total F-term scalar potential $V_F$ into two parts $V_{\rm KKLT}$ and $\Delta V$, where $\Delta V$ indicates the additional contribution due to the non-trivial dependence of the superpotential on the mobile D3 position.
As shown in \cite{Baumann1}, it is crucial to have such additional contribution for slow roll inflation, as it allows us to fine-tune the inflaton potential such that the slow-roll parameter $\eta$ can be made vanishingly small piece-wise. Most of the e-folds are then generated near the inflection point where the slow-roll parameter $\eta$ vanishes. 
Before moving on, let us close this section by noting that we can first stabilize the axion $\xi$ in the complex K\"ahler modulus $\rho$ at $\xi=0$ which corresponds to a minimum of the scalar potential in the $\xi$ direction as discussed in \cite{RaceTrack1}, unless $A(0)+B(0)<0$, $W_0<0$ and $a_1<a_2$ which yields a maximum at $\xi=0$.

\subsection{Stable Angular Trajectory}
\paragraph{}
The choice of D7 embedding in (\ref{Deff1f2})
is made not only because it is relatively symmetrical and remains supersymmetic without the presence of worldvolume flux, but more importantly, for the case of a single stack of D7-branes, it was shown to be the only known D7 embedding with  an angular stable trajectory.
In \cite{CGS1}, such trajectory has been extended to the entire deformed conifold, and is given by:
\begin{equation}
z_1=-\epsilon\cosh\frac{\tau}{2}\,,\label{z1stabletrajectory}
\end{equation}
The existence of such trajectory is crucial for obtaining a single field slow roll inflation model\footnote{Notice that along (\ref{z1stabletrajectory}), there remains a $SO(2)$ residual symmetry group, however such light angular degree of freedom has been shown to decouple from the canonical inflation \cite{CGS1}, and subsequently we shall not consider it in our discussion.}
As shown in detail in Appendix \ref{AngStabAnalysis}, the above trajectory remains angular stable when the second stack of D7s is introduced, as expected from symmetry argument. However to demonstrate the stability of the trajectory, it is crucial to simultaneously adjust the parameters specifying the potential such as $\mu_{1,2}$ and the deformation parameter $\epsilon$, hence their values
are constrained.
In the later section, we shall consider the situations where the mobile D3 brane moving in the region where $|\mu_2|^{2/3}< r <|\mu_1|^{2/3}$ as well as  $r <|\mu_{1,2}|^{2/3}$ and obtain the inflaton potentials. Having the F-term scalar potential valid for the entire deformed conifold throat allows us to consider different hierarchies between $\mu_1$ and $\mu_2$. For each case, we will numerically check the angular stability for the parameters yielding the inflationary trajectory.

Along the angular stable trajectory, the resultant scalar potential with the angular degrees of freedom integrated out is given by:
\begin{equation}
V_F(\tau,\sigma)=\frac{\kappa^2 S_0(\tau,\sigma)^2 }{3U(\tau,\sigma)^2}\left[U(\tau,\sigma)+6\left(\frac{W(\tau,\sigma)}{S_0(\tau,\sigma)}\right)
+6\Sigma(\tau,\sigma)\right]\label{VFtausigma}
\end{equation}
where
\begin{eqnarray}
\Sigma(\tau,\sigma)&=&(\epsilon^{4/3}\gamma)\left(K(\tau)\sinh\frac{\tau}{2}\right)^2
\left(K(\tau)\cosh\frac{\tau}{2}-\frac{S_1(\tau,\sigma)}{2 \epsilon^{1/3}\gamma  S_0(\tau,\sigma)}\right)^2\,\label{DefFtau}\\
K(\tau)&=&\frac{(\sinh 2\tau-2\tau)^{1/3}}{2^{1/3}\sinh\tau}\,.\label{DefKtau}
\end{eqnarray}
and $W(\tau,\sigma)\,,S_{0,1}(\tau,\sigma)$ denote functions $W(\tau,\sigma)$ and $S_{0,1}(z_1,\rho)$ defined in (\ref{DefW2stack}) and (\ref{DefSm},\ref{DefSm1}) with $z_1=-\epsilon\cosh\frac{\tau}{2}$ and $\rho=\sigma$ substituted.

We can now introduce the canonical inflaton field $\phi(\tau)$, which can be readily derived from the DBI action of a mobile $D3$ in the warped deformed conifold, with the explicit metric given by (\ref{ExpDeformMetric}):
\begin{equation}
\phi(\tau)=\sqrt{\frac{T_3}{6}}\int_\tau d\tau ' \frac{\epsilon^{2/3}}{K(\tau')}\,.\label{DefCanInflaton}
\end{equation}
One can see this definition has the following asymptotic limits:
\begin{eqnarray}
\tau\gg 1~&:&~\phi(\tau)\to\sqrt{\frac{3T_3}{2}} r\,,\label{largephi}\\
\tau\ll 1~&:&~\phi(\tau)\to\frac{\sqrt{T_3}}{2^{5/6}3^{1/6}}\epsilon^{2/3}\tau\,.\label{smallphi}
\end{eqnarray}
These limits would be useful when one tries to approximate the radial dependence of the volume modulus $\sigma$ near the tip region and for investigating the possible parameter choices for the inflationary trajectory at large radius.
One should also note that at generic radius, the canonical inflaton is only given by the integral expression (\ref{DefCanInflaton}), and conversely we should regard $\tau(\phi)$ as implicit function of the canonical inflaton.
For the calculations of slow roll parameters, however, the chain rule can be readily applied.

\subsection{Volume Stabilization and Single Field Inflation}
\paragraph{}
To study  volume stabilization and hence obtain the effective single field inflaton potential, one should also include the $\rm D3-\overline{\rm D3}$ potential
\begin{equation}
V_{D3\overline{D3}}(\tau,\sigma)=\frac{D_0}{U(\tau,\sigma)^2}\left(1-\frac{3D_0}{16\pi^2T_3^2{|\bf{y}-\bar{\bf{y}}|}^4}\right)\,.\label{VD3bD3}
\end{equation}
Here $D_0=2T_3a_0^4$ and $a_0=e^{-2\pi K/3g_sM}$ is the warp factor at the tip of the deformed conifold where the $\overline{\rm D3}$ is located; $\sepy$ is the separation between the D3 and
the $\overline{\rm D3}$ branes.
The leading term here gives a positive contribution to the total potential energy, and plays the crucial role of uplifting the KKLT type  AdS minimum to a dS one \cite{KKLT}. The remaining contribution in (\ref{VD3bD3}) corresponds to the warped $\rm D3-\overline{\rm D3}$ Coulombic attraction.
There can also be further distant uplifting source, e.g. $\overline{\rm D3}$ contributing to the vacuum energy or D7 branes carrying SUSY-breaking fluxes \cite{BKQ}. In the large volume limit, the most dominant term can be encoded as\footnote{The precise $U(\tau, \sigma)$ dependence in fact varies for different distant SUSY breaking sources, for $\overline{\rm D3}$ the potential $V_{\rm other} \sim U(\tau,\sigma)^{-2}$ and for D-term uplifting \cite{BKQ} induced by  D7-branes carrying SUSY breaking flux, $V_{\rm other} \sim U(\tau,\sigma)^{-3}$.
For concreteness, we consider the former case
which is the most dominant contribution
in the limit $U(\tau,\sigma)\gg 1$.} 
\begin{equation}
V_{\rm other}=\frac{D_{\rm other}}{U(\tau,\sigma)^2}\,,\label{Vother}
\end{equation}
such potential plays an important role in giving a small positive cosmological constant at the end of inflation after the $\rm D3-\overline{\rm D3}$ brane annihilation \cite{CGS1}.

We now derive the radial dependence of the stabilized volume $\sigma_\star(\phi)$ from the two field potential
\begin{equation}
{\mathbb{V}}(\tau,\sigma)=V_F(\tau,\sigma)+V_{D3\overline{D3}}(\tau,\sigma)+V_{\rm other}(\tau,\sigma)
\end{equation}
which satisfies the equation:
\begin{equation}
\frac{\partial{\mathbb{V}}(\tau,\sigma)}{\partial\sigma}\vline_{\sigma_\star}=0\,.\label{Defsigmastar}
\end{equation}
This is the so-called ``adiabatic approximation'', and the basic assumption here is that at every given point along the radial direction, the effective mass for the volume modulus $\sigma$ is always greater than that for the radial direction, even though the mass hireachy is not as large compared with the broken angular isometry direction such that it is stabilized at a constant value, but rather it is stabilized at its instantaneous minimum in the $\tau-\sigma$ plane.
Allowing for the volume modulus $\sigma(\tau)$ to vary with the radial coordinate is important to ensure angular stability and to obtain single field inflation.

Due to the exponential dependence of the ${\mathbb{V}}(\tau,\sigma)$ on $\sigma$,
the equation (\ref{Defsigmastar}) is a transcendental equation usually solved numerically. However, at least for $n_1=n_2$ an approximate analytic approach was adopted in \cite{Baumann1}, and it allows to obtain a qualitative understanding of the resultant potential. Assuming that in the large volume limit $\sigma\gg 1$, which can be ensured by having a large hierachy between $|S_0(0)|$ and $|W_0|$, one can set the $\sigma$ dependence in $U(\tau,\sigma)$ to a fixed value $\sigma_0$, where $\sigma_0$ is the particular solution to (\ref{Defsigmastar}) with $\tau=\phi=0$. Within such an approximate analytic approach, equation (\ref{Defsigmastar}) becomes a quadratic equation of the variable $X_{\star}=\exp(a\sigma_\star(\phi))$. To include the dependence on $\phi(\tau)$, we note that when $\sigma_0$ is large, the $\phi$ dependence only gives a small change in the stabilized volume, and we can perform a double expansion in $1/\sigma_0$ and small $\phi(\tau)$.
After solving the quadratic equation for $X_\star(\tau)$, one can deduce that for $n_1=n_2=n$,
\begin{eqnarray}
a\sigma_\star(\phi)&\approx&a\sigma_0\left(1+b_2\left(\frac{\phi}{M_p}\right)^2\right)\,,\label{Defasigmastar}\\
b_2&=&\frac{1}{(a\sigma_F)}\left(\frac{1}{3}+\frac{\Gamma}{8}\right)
+\frac{1}{(a\sigma_F)^2}\left(\frac{(4s-7)}{6}-\frac{\Gamma}{2}\right)+{\mathcal{O}}(1/\sigma_F^3)\,,\\
\Gamma&=&\left(\frac{2}{3}\right)^{2/3}
\frac{A_0\alpha_1(1+\alpha_1)^{1/n-1}+B_0\alpha_2(1+\alpha_2)^{1/n-1}}
{(n\beta^2)(A_0(1+\alpha_1)^{1/n}+B_0(1+\alpha_2)^{1/n})}\,.
\end{eqnarray}
Here, $\sigma_F$ is the solution to $D_{\sigma}W=\partial_{\sigma} V_F(\tau,\sigma)=0$ at the tip of the deformed conifold,
and the parameter $s$ is the uplifting ratio given by $s=\frac{(D_0+D_{\rm others})/U^2(\sigma_F,0)}{V_{F}(\sigma_F,0)}$.
In (\ref{Defasigmastar}) we have also introduced the following dimensionless parameters:
\begin{equation}
\alpha_{1,2}=\frac{\epsilon}{\mu_{1,2}}\,,~~~\beta=\sqrt{\frac{T_3}{6}}\frac{\epsilon^{2/3}}{M_p}\,,\label{Defalpha12beta}
\end{equation}
which will be useful later for finding workable parameter sets for the inflationary potential. Geometrically, $\alpha_{1,2}$ measure the depth of each stack of D7-branes within the deformed conifold region, so that $\alpha_{1,2}\to 1$ corresponds to sending the D7s into the highly warped region.
On the other hand, $\beta$ is proportional to the warped factor at the tip $\sim e^{-2\pi K/3g_sM}$ which can be deduced from the relation between four dimensional Planck mass $M_p$ and the warped compact volume.

Such semi-analytic expression for the stabilized volume obtained in the near tip region is expected to break down at large radius, including the definition of the canonical inflaton. It would be interesting to consider the resultant inflaton potentials derived from the semi-analytic (\ref{Defasigmastar}) and the numerical solutions to (\ref{Defsigmastar}), and to see how they differ in the CMB and other inflationary predictions.  
The semi-analytic expression is expected to capture the qualitative feature within its regime of validity, for our case the resultant single field inflaton is then given by:
{\small \begin{eqnarray}
{\mathbb{V}}(\tau(\phi),\sigma_\star(\phi))&=&\frac{\kappa^2 S_0(\tau,\sigma_\star(\tau))^2}{3U(\tau,\sigma_\star(\tau))^2}\left[U(\tau,\sigma_\star(\phi))
+6\left(\frac{W(\tau,\sigma_\star(\tau)}{S_0(\tau,\sigma_\star(\tau))}\right)
+6\Sigma(\tau,\sigma_\star(\tau))\right]+\frac{D(\tau)}{U^2(\tau,\sigma_\star(\tau))}\,.\nonumber\label{TotalInflatonpotential}\\
\\
D(\tau)&=&D_0\left(1-\frac{3D_0}{16\pi^2T_3^2{|\bf{y}-\bar{\bf{y}}|}^4}\right)+D_{\rm others}\,.\label{DefDphi}
\end{eqnarray}}
$\!\!$This potential will allow us to
demonstrate,
modulo microscopic compactification constraints,
that sufficient number of e-folds can be generated by tuning the parameters involved.

\section{On Gravitino Mass and the Inflation Scale}\label{KLbounddiscussion}
\paragraph{}
Before presenting the explicit parameter sets generating slow-roll inflation, we would like to pause here to discuss the issue of gravitino mass and the inflation scale raised in \cite{KLbound}, and consider the possiblity of obtaining a small gravitino mass using our superpotential (\ref{DefW2stack}).

At a supersymmetric minimum of an $\cN=1$ SUGRA F-term scalar potential, $D_\Sigma W=0$ for all moduli, and thus we naturally have an AdS vacuum with a negative cosmological constant given by $-3e^{\mathcal{K}/M_p^2}\frac{|W(\sigma_F)|^2}{M_p^4}$\footnote{Here for the clarity we have restore the relavant four dimensional Planck Mass $M_p$.}. In the KKLT scenario \cite{KKLT}, a dS vacuum is then constructed by introducing extra uplifting terms via an $\overline{\rm D3}$-brane at the tip of the warped throat or other distant sources, which retains the shape of the potential barrier to avoid decompactification. The uplifting only shifts the stabilized volume $\sigma_F$ by a small amount to $\sigma_0\sim\sigma_F$, and the gravitino mass in the uplifted vacuum is then given by $m_{3/2}^2(\sigma_0)=  e^{\mathcal{K}/M_p^2}\frac{|W(\sigma_0)|^2}{M_p^4}\approx e^{\mathcal{K}/M_p^2}\frac{|W(\sigma_F)|^2}{M_p^4}$. Therefore the gravitino mass is tied to the depth of the original AdS minimum $|V_{\rm AdS}|/M_p^2$.

As pointed out in \cite{KLbound}, in the simplest case where the non-perturbative superpotential is only generated by a single stack of D7 branes (or Euclidean D3 branes), such that $W(z,\rho)=W_0+A(z)e^{-a\rho}$; the Hubble scale for the various inflationary models constructed from the resultant uplifted de Sitter vacuum will obey the following bound
\begin{equation}
H\apprle m_{3/2}(\sigma_0)\,.\label{expKLbound}
\end{equation}
One can of course raise the scale of inflation, however this would distort the potential barrier and lead to runaway decompactification. This bound (\ref{expKLbound}) suggests that there
some tension between high scale inflation
 (favored by the observed amplitude of the power spectrum)
and low scale supersymmetry breaking
 (motivated by the hierarchy problem).

To resolve such issue, it was shown by generalizing to more than
a single non-perturbative instanton superpotential, there exists additional supersymmetric Minkowski vacua given by  \cite{KLbound,KLbound2}:
\begin{equation}
W(\sigma_{\rm Mink.})= 0\,,~~~\partial_\sigma W\vline_{\sigma_{\rm Mink.}}= 0\,,\label{MinkCond}
\end{equation}
such that in this vacuum, the gravitino mass can be made vanishingly small.
This is a special solution to $D_\sigma W=0$, hence still a minimum of the scalar potential $V_F$ in the $\sigma$ direction. Notice, however, additional conditions need to be imposed on
the parameters in the superpotential (\ref{DefW2stack}), e.g., $W_0$, $A_0$, $B_0$ $n_1$ and $n_2$, for such vacuum to exist.
Importantly, in such vacuum, the gravitino mass is
no longer tied to the height of the potential barrier.
Therefore, inflation can proceed at a high scale without pushing the gravitino mass outside the
reach of current and upcoming accelerators.

A remaining question is whether one can construct a viable inflation model
in this scenario
 while evading the bound in \cite{KLbound}. Previous attempts \cite{ CKLQ,KL2, AccidentalInflation, BO1, BO2,Abe:2008xu}
have focused on using K\"ahler modulus as the inflaton.
Here, the inflaton is an open string mode.
As we shall see, having the inflaton in the open string sector allows more freedom in model building.
Given that the gravitino mass is disentangled from the magnitude of the potential barrier stabilizing the minimum, the additional energy from
the open string mode
may allow inflation to take place at  a higher energy scale than the gravitino mass, without leading to runaway decompactification. In the following, we shall demonstrate that slow-roll inflaton potential can be constructed with appropriate choices of microscopic parameters and that the Hubble scale can exceed the gravitino mass in the uplifted vacuum.

\section{Sample Parameters and Inflationary Trajectories}\label{Results}
\paragraph{}
In this section we present some explicit examples of the parameter sets which allow for a sufficient number of e-folds to be generated. We shall consider three different cases characterized by different relative values of
the embedding parameters $\mu_1$ and $\mu_2$ and the
prefactors\footnote{For a Euclidean D3-brane, the prefactor represents a one-loop determinant of fluctuations around the instanton. For D7-branes, the prefactor comes from a threshold correction to the gauge coupling on the D7-branes.}
$A_0$ and $B_0$.
In the first two cases, the mobile D3 brane has different relative positions to the moduli-stabilizing D7s. Without loss of generalities, we shall consider the simplified situation with $n_1=n_2=n$, and inflation takes place after uplifting an AdS vacuum.
In the third case, we shall consider inflation upon uplifting a Minkowski vacuum where it is therefore crucial to have $n_1\neq n_2$.

For a sustained period of inflation, the slow-roll parameters $\varepsilon, \eta$ given by
\begin{equation}
\varepsilon(\phi) \equiv \frac{1}{2}\left(\frac{1}{\mathbb{V}(\phi)}\frac{d{\mathbb V(\phi)}}{d\phi}\right)^2,\qquad \eta(\phi) \equiv \frac{1}{\mathbb{V}(\phi)}\frac{d^2{\mathbb V}(\phi)}{d\phi^2},\label{DefSlowRollParas}
\end{equation}
need to remain small {\it i.e.} $\varepsilon, |\eta|\ll 1$.
The number of e-folds
% at by the inflaton at some value and
before
the end of inflation is then given by
\begin{equation}
N(\phi)=\int_{\phi_{\rm e}}^{\phi} \frac{d\phi'}{\sqrt{2 \varepsilon(\phi')}}.\label{DefN}
\end{equation}
Here $\phi_{\rm e}$ denotes the value of $\phi$ when inflation ends, that corresponds to $\varepsilon(\phi_{\rm e})=1$.
In each of the three cases, sufficient number of e-folds $(\apprge 60)$ relies
% crucially
on the existence of %the so-called
 an
 ``inflection point'' in ${\mathbb{V}}(\phi)$, where $\eta=0$ and in its vicinity $\varepsilon\ll 1$. We shall demonstrate such point can be obtained with judicial choices of the allowed microscopic parameters, and we calculate (\ref{DefN}) numerically generated around
 the inflection point
  since this region is where the majority of e-folds are expected to be generated.
It is
important to note though that it is
 possible to calculate the total number of e-folds using our potential with appropriate initial and
 final
  conditions, since it is valid in the entire deformed conifold including the tip region where the branes annihilate
   ending inflation.
For each set of parameters used, we also check the angular stability of the resultant trajectory in Appendix \ref{AngStabAnalysis}.
Notice that the inflection point inflation is known to suffer from the ``overshoot problem'' \cite{Overshoot1,Underwood:2008dh}, such that the initial field position needs to be near the inflection point itself to generate large number of e-folds.
Here we refrain from discussing further this issue,
and simply refer the readers to
\cite{Overshoot1,Underwood:2008dh}.
In obtaining the numerical number of e-folds, we shall simply assume that inflaton indeed rolls
through the inflection point and perform the integration using (\ref{DefN}) with upper and lower limits as shown in the various figures.

%%%%%%%%%%%%%%%%%%%%%%%%%%%%%%%%%%%%%%%%%%%%%%%%%%%%%%%%%%%%%%%%%%%%%%%%%%%%%%%
\subsection{Case A: $0<r<|\mu_2|^{2/3}<|\mu_1|^{2/3}$}\label{CaseA}
\paragraph{}
In this simplest case, both stacks of D7 branes are far from the tip region and inflation takes place at $r< |\mu_{1,2}|^{2/3}$.
This case is very similar to the configuration studied in \cite{Baumann1}. Notice however it is still necessary for the two stack of D7 branes to be seperated at a distance larger than the local string length to
avoid the presence of light $7$-$7$ open string modes (which can lead to
the disappearance of the superpotential (\ref{DefW2stack})).
While inflation is UV sensitive, it is not very sensitive to the addition of extra D7-branes at large radial distance.
A sample set of parameters is listed below. The potential becomes very flat in the region $0.072<\phi/M_p<0.148$, near the inflection point $\phi/M_p\approx 0.1105$. The number of e-folds generated in this region
is around $N \approx 107$.
\begin{center}
\begin{table}[htp]\label{TableA}
\begin{tabular}{|c|c|c|c|c|c|c|c|c|c|c|} \hline
$n$& $A_0$ & $B_0$ & $\alpha_1$ & $\alpha_2$& $\beta $& $W_0$ &$ s$ &$ p$ &$\sigma_F$&$m_{3/2}(\sigma_0)$ \\ \hline
8& 1 & 0.95 & 1/100 & 1/65 &1/309& $3.349 \times 10^{-4}$&1.05 & 0.9 & 13.6799 & $2.10456\times 10^{-6}$\\\hline
\end{tabular}
\caption{Compactification data for Case A}
\end{table}
\end{center}
\begin{figure}[ht]
\begin{center}
%\includegraphics[
%width=0.4\textwidth]{CaseAglobpot.eps}
\includegraphics[
width=0.4\textwidth]{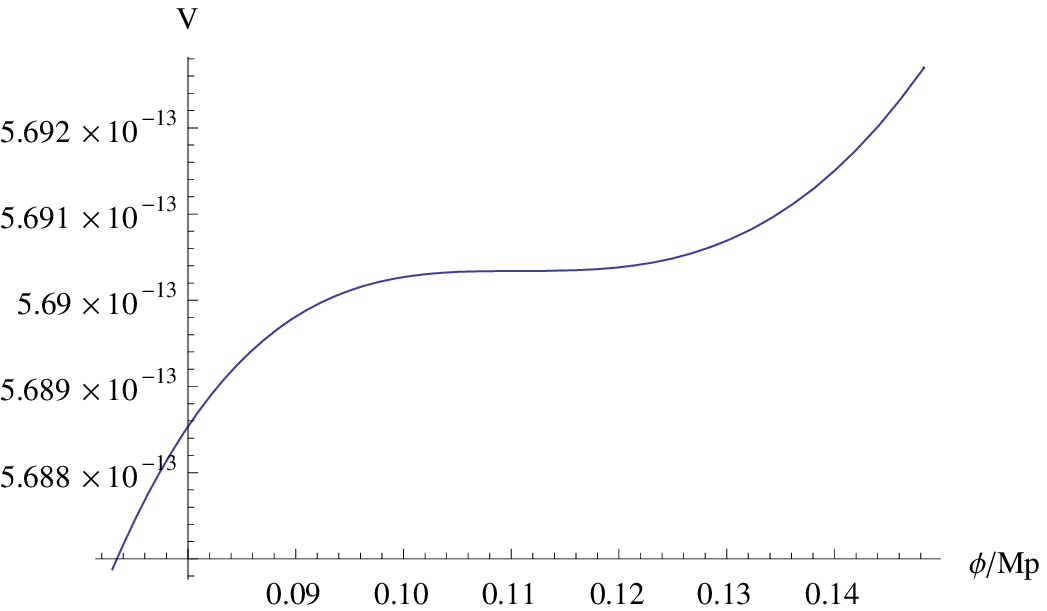}
\includegraphics[
width=0.4\textwidth]{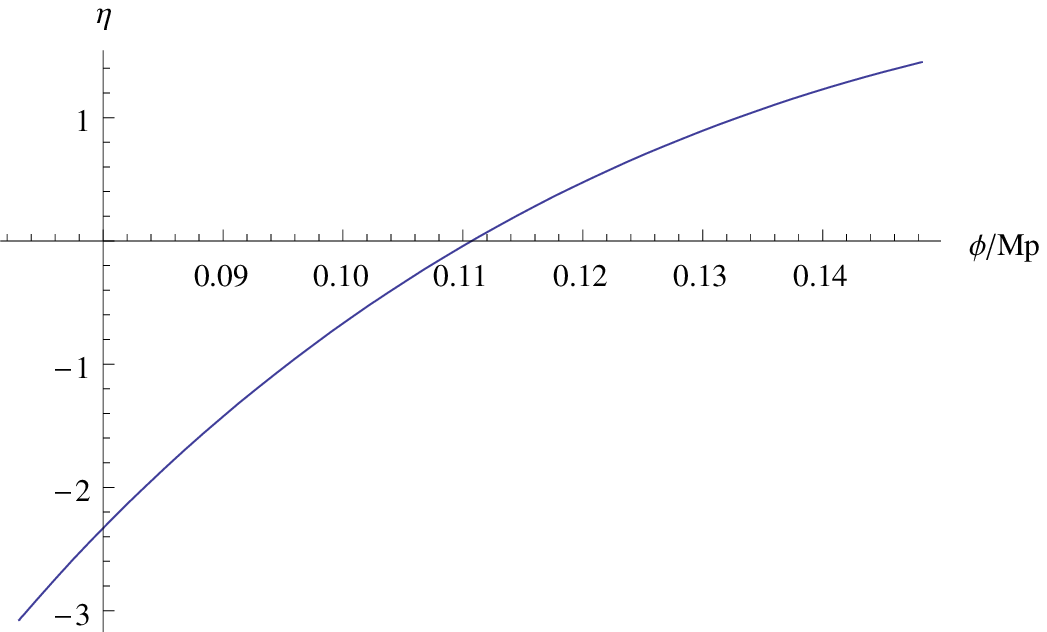}
\caption{Case A: The inflaton potential ${\mathbb{V}}(\phi)$ in Planck units and the slow-roll parameter $\eta$ near the inflection point.}
%\caption{The Slow-Roll Parameter $\eta$ near the inflection point.}
\end{center}
\end{figure}
In Table 1, $A_0$, $B_0$ and $W_0$ have mass dimension three and are expressed in Planck units; $m_{3/2}(\sigma_0)$ is the gravitino mass in Planck unit in the resultant uplifted dS vacuum; %whereas
 the parameter $p$ is defined such that $\frac{D_{\rm others}}{D_0}=\frac{p}{(1-p)}$. It is interesting to note that the Hubble scale in this case is clearly less than the gravitino mass, as expected from
 the uplifting of a generic supersymmetric AdS minimum \cite{KLbound}.
%%%%%%%%%%%%%%%%%%%%%%%%%%%%%%%%%%%%%%%%%%%%%%%%%%%%%%%%%%%%%%%%%%%%%%%%%%%%%%

%%%%%%%%%%%%%%%%%%%%%%%%%%%%%%%%%%%%%%%%%%%%%%%%%%%%%%%%%%%%%%%%%%%%%%%%%%%%%%%
%\newpage
\subsection{Case B: $0<|\mu_2|^{2/3} <r <|\mu_1|^{2/3}$}\label{CaseB}
\paragraph{}
This case is different from the previously considered configuration in \cite{Baumann1}, where we have a large hierarchy between the embedding parameters of the two D7-brane stacks, and most of the e-folds are generated as the mobile D3-brane moves between them.
For the set of parameters listed below, the near flat region in ${\mathbb{V}}(\phi)$ is around $0.072<\phi/M_p<0.145$ with the inflection point located around $\phi/M_p\approx 0.11$. The number of e-folds generated in that region is $N \approx 146$. Again, the Hubble scale is less than the gravitino mass
%as the generic feature of uplifted AdS vacuum in the simplest KKLT scenario.
which is a generic feature in uplifting an AdS vacuum in the simplest KKLT scenario \cite{KLbound}.
\begin{center}
\begin{table}[htp]
\begin{tabular}{|c|c|c|c|c|c|c|c|c|c|c|} \hline
$n$& $A_0$ & $B_0$ & $\alpha_1$ & $\alpha_2$& $\beta $& $W_0$ &$ s$ &$ p$&$\sigma_F$&$m_{3/2}(\sigma_0)$ \\ \hline
8& 1 & 0.001 & 1/100 & 1/3 &1/315& $3.349 \times 10^{-4}$&1.065 & 0.9 &12.7524 &$2.31773\times 10^{-6}$\\\hline
\end{tabular}
\caption{Compactification data for Case B}
\end{table}
\end{center}
\begin{figure}[htp]
\centering
%\includegraphics[
%width=0.4\textwidth]{CaseBglobpot}
\includegraphics[
width=0.4\textwidth]{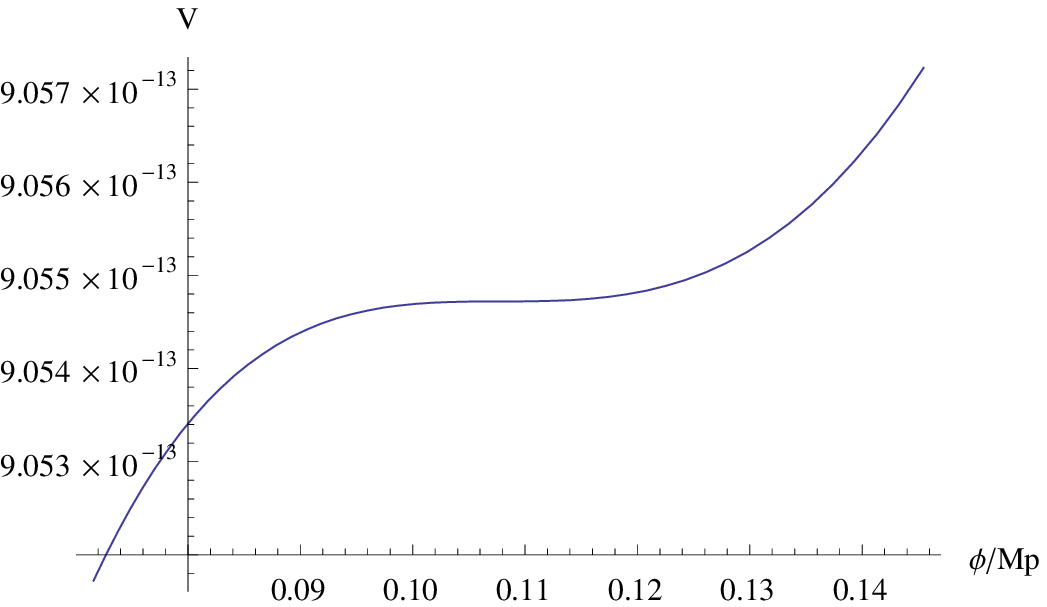}
\includegraphics[
width=0.4\textwidth]{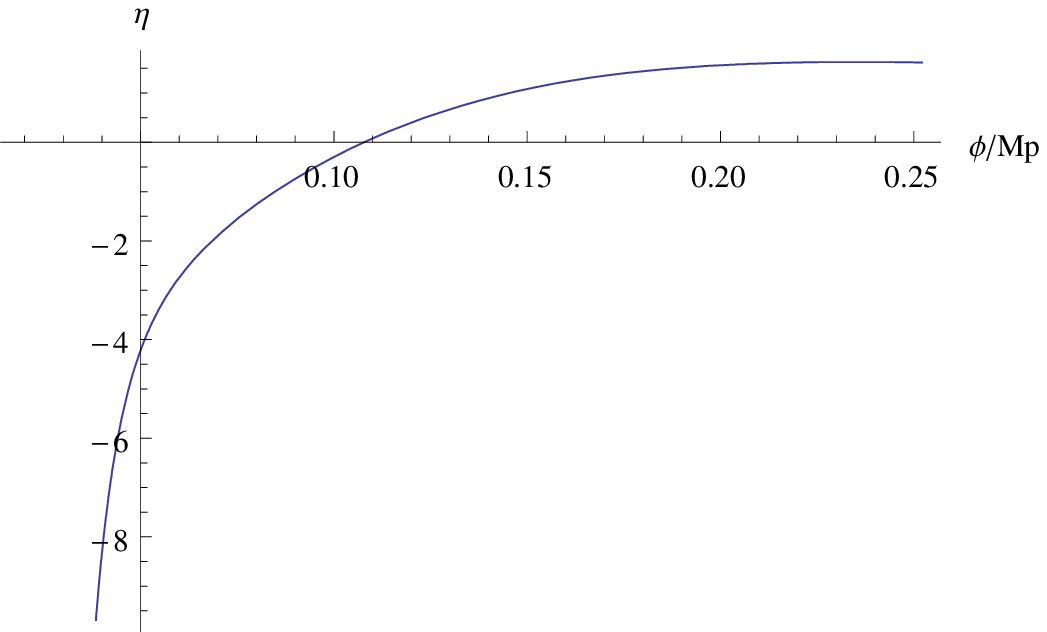}
\caption{Case B: The inflaton potential ${\mathbb{V}}(\phi)$ in Planck units and the slow-roll parameter $\eta$ near the inflection point.}
\end{figure}
Notice that in Table 2, we have performed a $10^{-3}$ tuning between $A_0$ and $B_0$ to obtain an inflection point inflaton potential. Without such tuning, i.e., $|A_0|\sim |B_0|$, due to cancellation of the forces, one would instead obtain
generically a metastable local de Sitter minimum which localizes the mobile D3 brane at some intermediate radius within the deformed conifold. From this perspective, inflation can be regarded as accidental in such two stack configurations, resulting from
additional tuning of an otherwise metastable minimum.
%%%%%%%%%%%%%%%%%%%%%%%%%%%%%%%%%%%%%%%%%%%%%%%%%%%%%%%%%%%%%%%%%%%%%%%%%%%%%%
\subsection{Case C: $A_0 B_0 < 0$}\label{CaseC}
\paragraph{}
This is arguably the most interesting case. Instead of beginning with an AdS vacuum, we first
fine-tune $W_0$ to obtain an additional supersymmetric Minkowski vacuum at $\sigma=\sigma_{\rm Mink}$  as in \cite{KLbound}.
{\small
\begin{eqnarray}
\sigma_{\rm Mink} &=&\frac{1}{a_1-a_2}\log\left|-\frac{A_0}{B_0}\frac{a_1(1+\alpha_1)^{\frac{1}{n_1}}}{a_2(1+\alpha_2)^{\frac{1}{n_2}}}\right|\,,\label{DefsigmaMink}\\
 W_0 &=& -A_0(1+\alpha_1)^{\frac{1}{n_1}}\left|-\frac{A_0}{B_0}\frac{a_1(1+\alpha_1)^{\frac{1}{n_1}}}{a_2(1+\alpha_2)^{\frac{1}{n_2}}}\right|^{-\frac{a_1}{a_1-a_2}}
-B_0(1+\alpha_2)^{\frac{1}{n_2}}\left|-\frac{A_0}{B_0}\frac{a_1(1+\alpha_1)^{\frac{1}{n_1}}}{a_2(1+\alpha_2)^{\frac{1}{n_2}}}\right|^{-\frac{a_2}{a_1-a_2}}\,.\label{W0Mink}
\end{eqnarray}}
$\!\!$Notice that without such fine-tuning, we would only have a single AdS vacuum in the F-term scalar potential, and the scale of SUSY breaking is large and tied to the barrier height. In the Minkowski vacuum, however, the scale of SUSY breaking is manifestly zero. When a suitable extra uplifting terms are included\footnote{Notice that the uplifting ratio ``s'' ceases to be meaningful in this case, as $V_F(\sigma_{\rm Mink})=0$.}, the local metastable minimum still exists to avoid runaway decompactification. Furthermore, even when the radial dependence is included, it only causes a small shift in $\sigma$, hence the resultant gravitino mass remains small as the moblie D3 brane moves down the throat.
\begin{center}
\begin{table}[htp]
{\small
\begin{tabular}{|c|c|c|c|c|c|c|c|c|c|c|} \hline
$n_1$&$n_2$& $A_0$ & $B_0$ & $\alpha_1$ & $\alpha_2$& $\beta $& $W_0$ &$ p$&$\sigma_{\rm Mink}$&$m_{3/2}(\sigma_0)$ \\ \hline
38&40& 1 & -0.9 & 1/80 & 1/75 &10/2993& $-2.29481 \times 10^{-3}$ & 0.9 &18.948 &$3.29719\times 10^{-8}$\\\hline
\end{tabular}}
\caption{Compactification data for Case C}
\end{table}
\end{center}%%%%%%%%%%%%%%%%%%%%%%%%%%%%%%%%%%%%%%%%%%%%%%%%%%%%%%%%%%%
\begin{figure}[htp]\label{CaseCscalarpotential}
\centering
\includegraphics[
width=0.4\textwidth]{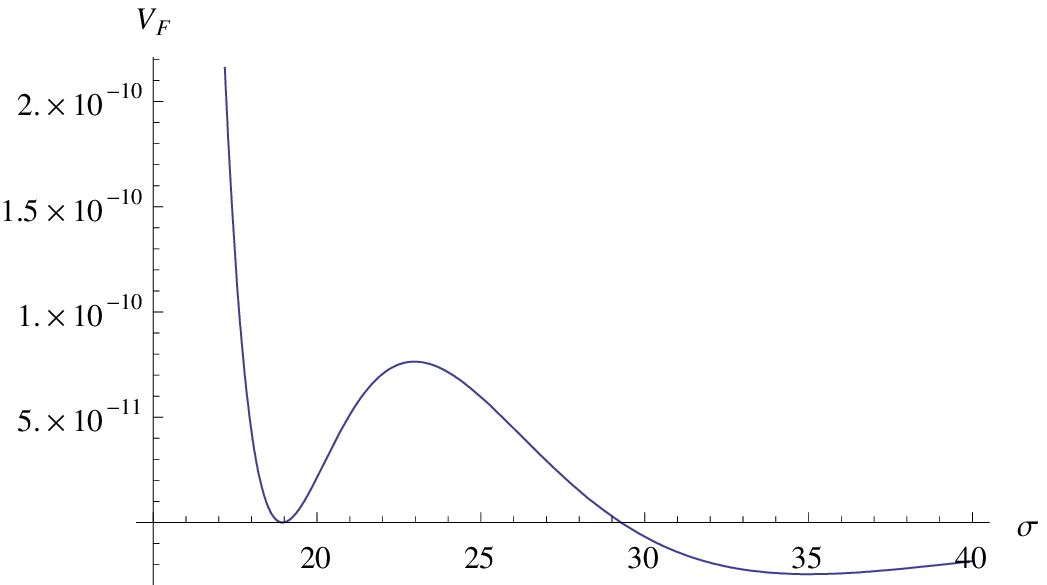}
\includegraphics[
width=0.4\textwidth]{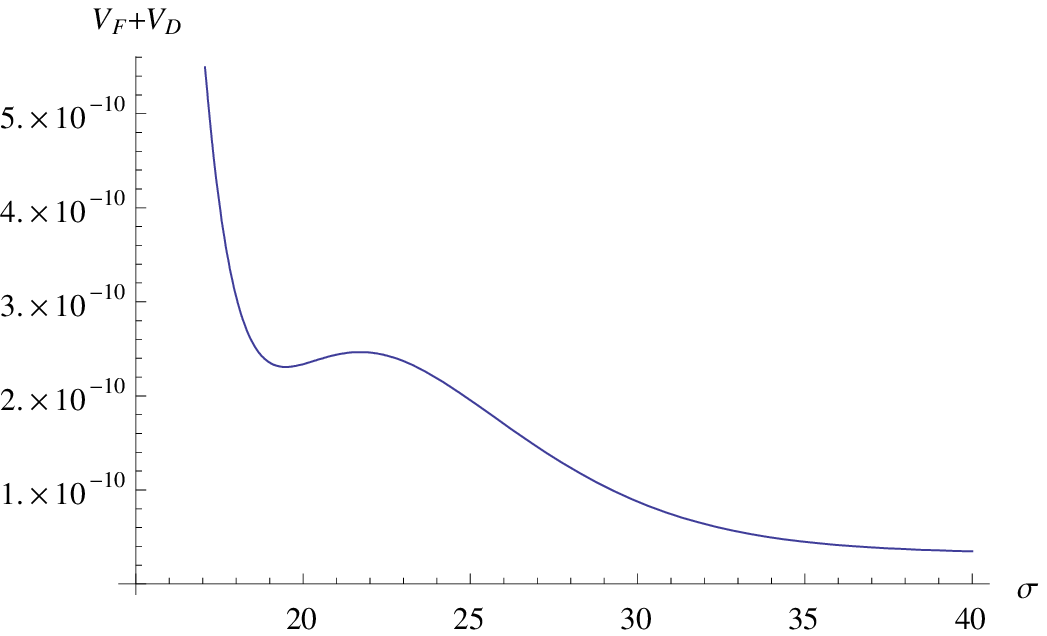}
\caption{Case C: The Scalar Potential at $\tau=0$ before and after uplifting}
%\end{figure}
%%%%%%%%%%%%%%%%%%%%%%%%%%%%%%%%%%%%%%%%%%%%%%%%%%%%%%%%%%%%%%%%
%\begin{figure}[]
%\centering
%\includegraphics[
%width=0.4\textwidth]{CaseCglobpot}
\includegraphics[
width=0.4\textwidth]{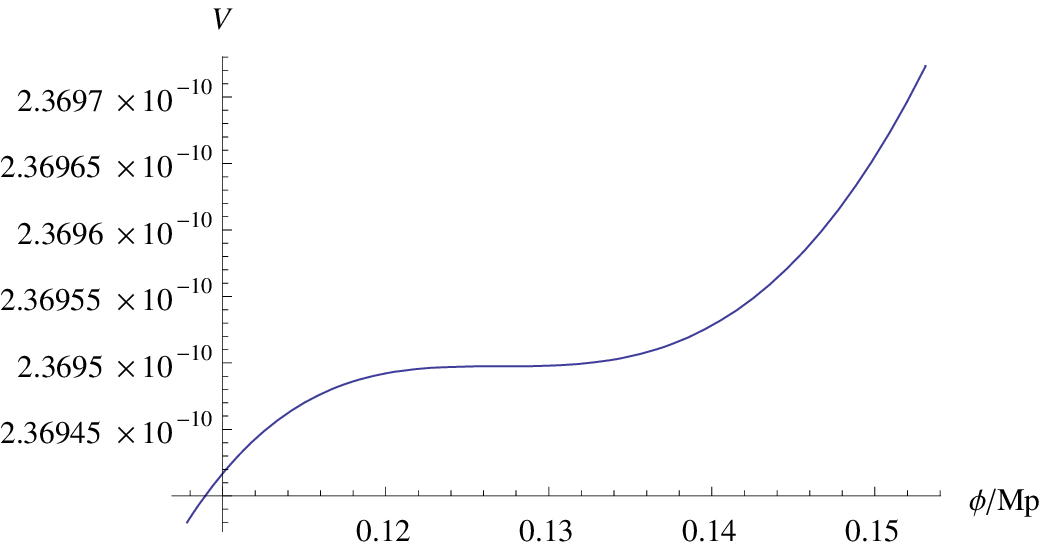}
\includegraphics[
width=0.4\textwidth]{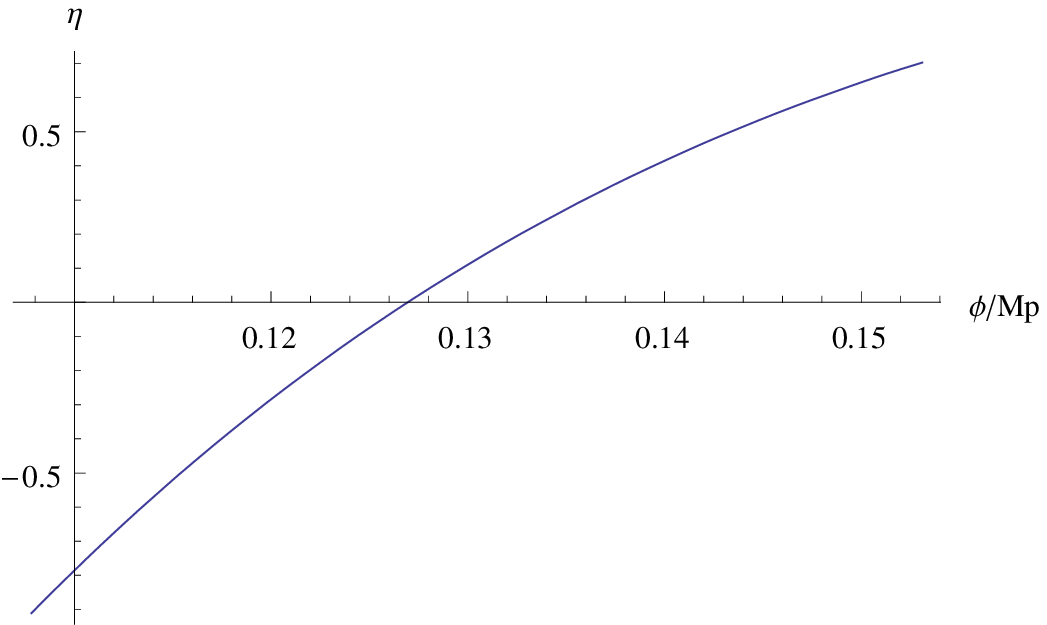}
\caption{Case C: The Inflaton Potential ${\mathbb{V}}(\phi)$ in Planck Unit and slow-roll parameter $\eta$ near the inflection point.}
\end{figure}
%%%%%%%%%%%%%%%%%%%%%%%%%%%%%%%%%%%%%%%%%%%%%%%%%%%%%%%%%%%%%%%%%%%%%%%%%%%%%%%
For the set of parameters listed above, the inflection point is near $\phi/M_p \approx 0.127$, and around $257$ e-folds are generated in the region of field space where $0.12 < \phi/M_p < 0.14$. Most importantly, the Hubble scale in this case is $\approx 8.89\times 10^{-6}$ in Planck units, which is about 270 times larger than the gravitino mass $m_{3/2}(\sigma_0)$.
While this provides a proof of concept that the Hubble scale during inflation can exceed the gravitino mass,
further parameter scanning is needed to show that the observed amplitude of the power spectrum and TeV scale soft masses can be simultaneously obtained. Such analysis also allows us to quantify the degree of fine-tuning involved, for some related recent work, see \cite{Hoi:2008gc, CG1}.
It is interesting to compare inflation on an ``open racetrack"  with the  racetrack inflationary scenario in \cite{RaceTrack1} where the Kahler modulus (more precisely, the axion component) which appears in the non-perturbative superpotential
also plays the role of the inflaton.
Clearly, if the inflaton is an additional field (here, an open string mode) other than the modulus being stabilized by the racetrack potential, one finds more flexibilities in model building.
For example, it has been argued that in racetrack inflation, additional non-perturbative terms
beyond the minimal racetrack potential
seems necessary for  inflation to occur \cite{BO1}.
This is not the case for the ``open racetrack".
Furthermore, inflation is sensitive to dimension six, Planck suppressed corrections to the inflaton potential.
Therefore, a global shift symmetry was invoked in racetrack inflation
to protect the inflaton potential from further UV corrections.
Here, no such symmetry is imposed and
because a lot is known about the local geometry of the throat, not only can we
compute the renormalizable part of the inflaton potential,
 we can apply gauge/string duality  to characterize such corrections due to bulk physics \cite{HoloBulkeffects}.

%\newpage
\section{Discussions}\label{Discussions}
\paragraph{}
In this paper, we present a 
variant of warped D-brane inflation 
 by
 introducing multiple stacks of moduli-stabilizing D7-branes in a warped
 throat.
We used the warped deformed conifold as an illustrative example, though the open racetrack inflationary scenario introduced here can be adopted to more general backgrounds.
We have considered various
 configurations of D7-branes with different relative positions with respect to the mobile D3-brane, leading to qualitatively distinct inflationary scenarios.
 Furthermore using the racetrack-like superpotential, we demonstrated that a supersymmetric Minkowski vacuum can be obtained  prior to uplifting and constructed an explicit inflationary model where the Hubble scale 
 exceeds the gravitino mass. Thus, the phenomenological bound on the inflation scale, i.e.,  $H \apprle m_{3/2}(\sigma_0)$  as pointed out by \cite{KLbound} can be evaded (albeit with fine-tuning), and to the best of our knowledge, this is the first realization of such models in the context of brane inflation.

There are a number of interesting directions one can further pursue.
A natural direction is  to generalize the analysis here to configurations involving other D7-brane embeddings and to construct explicit models. An obvious obstacle to be overcome, however, 
is to obtain an angular stable trajectory.
This can be difficult in the intermediate region; but if there is a large separation between different stacks of D7-branes, we expect there exists region in between the D7-branes where single field inflation
can be realized.
It is important to note however that the requirement of an angular stable trajectory is imposed
only for calculational convenience as the resulting dynamics of an effective single field model can be analyzed semi-analytically.
A priori, one can numerically solve for the inflationary trajectory of a multi-field model, and
in fact such multi-field models are interesting because their dynamics can 
lead to interesting non-Gaussian features in the CMB\footnote{For recent work on deriving the bispectrum of general multi-field inflation, extending the single field result of \cite{Chen:2006nt}, see \cite{Langlois:2008qf,Arroja:2008yy} (see also \cite{Gao:2008dt} for a restrictive case, and \cite{Naruko:2008sq} for
multi-brid inflation).}.
Besides the technical difficulties of finding angular stable trajectories, generic configurations of multiple D7-brane stacks can intersect along their worldvolume. 
The light open string modes localized at these intersections can modify the non-perturbative gaugino condensate terms in the superpotential. It is therefore 
 important to study these modifications   in order
 to realize these configurations concretely.

We have demonstrated as a proof of concept that high scale inflation and low scale supersymmetry can 
in principle coexist in the current setup.
However, whether there exists a set of microscopic parameters that give rise to 
predictions  compatible with cosmological data {\it and} at the same time having a supersymmetry breaking scale in the phenomenologically relevant range requires more extensive numerical work.
It would be interesting to perform a scan of the microscopic parameter space and compare the corresponding cosmological signatures with data. Such analysis may also quantify the amount of fine-tuning needed to evade the bound in \cite{KLbound}.
Of course, a quantitative study of soft supersymmetry breaking masses would require 
an embedding of the Standard Model and the supersymmetry (SUSY) 
breaking/mediation sector in warped 
compactifications.
Since the masses of the messenger fields depend on the separation between the Standard Model and the SUSY breaking branes (see, e.g., \cite{Diaconescu:2005pc}), 
the mechanism of stabilizing the separation between D-branes at a finite tunable distance 
discussed in this work may be relevant for embedding various 
supersymmetry breaking scenario in string theory.
We hope to return to all these interesting issues in the future.

\subsection*{Acknowledgements}
\paragraph{}
We are grateful to Konstantin Bobkov, Fang Chen,
Jim Cline, Shamit Kachru, Renata Kallosh, Andrei Linde, Yu Nakayama, Peter Ouyang, Fernando Quevedo, Stuart Raby, Alexander Westphal, and Piljin Yi for discussions.
The work of HYC and GS  was supported in part by NSF CAREER Award No. PHY-0348093, DOE grant DE-FG-02-95ER40896, a Research Innovation Award and a Cottrell Scholar Award from Research
Corporation, a Vilas Associate Award from the University of Wisconsin, and a John Simon Guggenheim Memorial Foundation Fellowship.
HYC and GS also thank the Stanford Institute for Theoretical Physics and SLAC for hospitality and support while this work was written.
LYH is supported by the Gates Cambridge Trust.

\appendix
\section{Gaugino Condensate in the Deformed Conifold}\label{GauginoCondensate}
\paragraph{}
Here we demonstrate that the D3-brane moduli dependence of the gaugino condensate is independent of the presence of additional D7 brane worldvolume flux.
We shall first discuss this from a closed string perspective following \cite{Gaugino1, Baumann1}, and then present a supporting open string calculation by
generalizing
the results in toroidal orientifolds \cite{Berg}
to include worldvolume flux.

\subsection{Closed String Perspective}\label{CloseString}
\paragraph{}
In the singular conifold limit,
the dependence of the D7 gaugino condensate on the position of mobile D3 was deduced in \cite{Gaugino1}. Essentially the calculation amounts to taking into account the perturbation of the D3-brane on the warp factor of the holomorphic four cycle wrapped by the D7-branes.
From the DBI action of D7-branes, this translates into a position dependent shift in the four dimensional Wilsonian coupling, which, by dimensional transmutation, becomes the resultant gaugino condensate.
As shown in \cite{Gaugino1},
 the total warped four cycle volume with the backreaction of the D3-branes
 can also be expressed as the real part of a holomorphic function, this in turns ensures the holomorphy of the D7 gauge coupling.

To generalize this result in the deformed conifold, one first notices that there is an extra bulk complex three form flux $G_3=F_3-\tau H_3$ also present  in this background \cite{KS}.
The supersymmetric condition on the D7 brane can be given in geometrical terms as
\begin{equation}
\hat{J}\wedge {\mathcal{F}}=0\,.\label{kappasymmetry}
\end{equation}
Here
$\hat{J}$ is the pull-back of the bulk K\"ahler form onto the D7 brane,
${\mathcal{F}}=\hat{B}_2+2\pi \alpha ' F_2$ is the gauge invariant 2-form,
where
$\hat{B}_2$ is the pullback of the supergravity NS-NS 2-form, and
$F_2$ is the gauge field strength on the worldvolume of the D7-branes.
The condition (\ref{kappasymmetry}) also needs to be supplemented with the Bianchi identity:
\begin{equation}
d\mathcal{F}=\hat{H}_3\,.\label{anomaly}
\end{equation}
While these conditions can be satisfied without the inclusion of worldvolume flux for the Kuperstein embedding, other choices of D7-brane embeddings in general require a non-trivial $F_2$,
 and in many cases, the solutions can only be constructed numerically \cite{COSD7}.
As shown in \cite{Baumann1}, if the supersymmetric condition (\ref{kappasymmetry}) is satisfied, the warped factor dependence on $\mathcal{F}$ canceled out  between the DBI and the Chern-Simons terms at the leading non-trivial order of the
small four dimensional gauge field strength expansion.
At this order, $\mathcal{F}$ only contributes a cut-off dependent shift to the four cycle volume, which is independent of the mobile D3 position, therefore $\mathcal{F}$ only
changes the overall magnitude of the gaugino condensate, not its dependence on the moduli. Such conclusion is consistent with the preservation of residual global symmetry and the holomorphy of the superpotential.

\subsection{Open String Perspective}\label{Openstring}
\paragraph{}
To support our discussion about the effect of the worldvolume flux on the D7 gaugino condensate, here we present a complementary open
string calculation in
simple toroidal orientifolds, generalizing \cite{Berg}.
As is well known, one loop open string amplitudes are related to closed string tree level exchanges between the boundary states representing the D-branes via a modular transformation of string world-sheet coordinates. We review here the tools needed for these computations.

The calculations are particularly simple if we use the light-cone gauge fixed GS boundary states. In this framework the $Dp$ branes are represented as Euclidean $p+1$ branes. In the light-cone gauge $p$ is limited to $-1<p<7$.
Consider boundary states corresponding to a $Dp$ brane. They satisfy the following boundary conditions
\begin{eqnarray}\label{bc}
\partial X^I - \mathcal{T}^p_{I J}\bar{\partial} X^{J}  |B_{I_1...I_{p+1}}, \zeta>  = 0,  \\
S^m + i\zeta \cM^p_{mn}\tilde{S}^n |B_{I_1...I_{p+1}}, \zeta> =  0, &\qquad& \tilde{S}^{\dot{m}} + i\zeta \cM^{p}_{\dot{m}\dot{n}}\tilde{S}^{\dot{n}} |B_{I_1...I_{p+1}}, \zeta> =0,  \label{superreflect}
\end{eqnarray}
where $\zeta = \pm 1$ for brane and anti-brane respectively. The matrices $\cM^{p}_{mn}$ and $\cM^{p}_{\dot{m}\dot{n}}$ determine the supersymmetry preserved by the boundary state and are given by \cite{Gutperle}
\begin{equation}
\cM^p_{mn} = (\gamma^1\gamma^2 ...\gamma^{p+1})_{mn}, \qquad   \cM^p_{\dot{m}\dot{n}} = (\gamma^1\gamma^2 ...\gamma^{p+1})_{\dot{m}\dot{n}}.
\end{equation}
These $\gamma$ matrices are building blocks of the eight-dimensional gamma matrices and are defined in \cite{GSW}.
The matrix $\mathcal{T}^p$ is an element of $SO(8)$ (i.e. rotations along the transverse directions ) and is expressible as
\begin{equation}
\mathcal{T}^p = \exp{(\Omega_{AB}\Sigma^{AB})},
\end{equation}
where $\Sigma^{AB}$ are $SO(8)$ generators in the vector representation.
For a (Euclidean) $Dp$-brane stretched along $\{x^1... x^{p+1}\}$ and in the absence of worldvolume gauge field strength,
\begin{equation}
\mathcal{T}^p = \left(\begin{array}{cc}
-I_{p+1} & 0 \\
0            &  I_{7-p}
\end{array}
\right),
\end{equation}
where $I_n$ denotes an $n \times n$ unit matrix.
which basically defines the Neumann boundary condition along the first $p+1$ directions and the Dirichlet boundary condition along the remaining $7-p$ directions.
The boundary state satisfying the boundary conditions above is then given by \cite{Gutperle}
\begin{equation}\label{boundarystate}
 |B_{I_1...I_{p+1}}, \zeta>  = \exp(\sum_{q>0}(\frac{1}{q}\mathcal{T}^p_{IJ}\alpha^I_{-q}\tilde{\alpha}^J_{-q})-i \cM^p_{mn}S^m_{-q}S^n_{-q}) |B_{0(I_1...I_{p+1})}, \zeta>
\end{equation}
where
\begin{equation}\label{zero_modes}
 |B_{0(I_1...I_{p+1})}, \zeta>  = T_{p} [ (\mathcal{T}^p_{IJ}|I>\otimes |J> + i\zeta \cM^p_{\dot{m}\dot{n}}|\dot{m}>\otimes |\dot{n}>)]_{p=0}.
\end{equation}
The normalisation $T_{p}$ is determined by comparing the tree level exchange between two parallel branes with the corresponding one-loop open string calculation, and can be identified with the
Dp-brane tension.
The effect of worldvolume gauge fields can be captured by turning on a flux condensate on the boundary state. This alters the boundary condition satisfied by the closed string at the boundary. The boundary conditions along the Neumann directions are then given by
\begin{equation}
\partial_{n}X^{I} + \mathcal{F}_{IJ}\partial_{t}X^{J} |B_{I_1...I_{p+1}}, \zeta,\mathcal{F}>  = 0,
\end{equation}
where $n$ and $t$ are the normal and tangential directions to the boundary respectively, and $\mathcal{F}$ is the generalized worldvolume flux defined earlier.
To rewrite it as in equation (\ref{bc}) above,
we have \cite{Gutperle}
\begin{equation}
\mathcal{T}^p_{IJ}(\mathcal{F}) = - (1- \mathcal{F})_{IK}(1+\mathcal{F})^{-1}_{KJ}.
\end{equation}
The form of the matrix $\cM^p$ is also altered since the supersymmetries preserved by the boundary state is dependent on the worldvolume flux. In fact we have \cite{Gutperle}
\begin{equation}
\cM^p(\mathcal{F}) = \prod^{(p+1)/2}_{i=1} \frac{1}{\sqrt{1+f_i^2}}(1+f_i\gamma^{2i-1} {}^{2i})\gamma^1 ... \gamma^{p+1},
\end{equation}
where $\mathcal{F}$
have been arranged in a block-diagonal form, with each block given by
\begin{displaymath}
\left(\begin{array}{cc}
0 & f_i \\
-f_i          & 0
\end{array}\right).
\end{displaymath}	
The normalization in the presence of flux $\mathcal{F}$ is given by
\begin{equation}
\tilde{T}_{p}(\mathcal{F}) = T_{p}\sqrt{\det(1+\mathcal{F})}.
\end{equation}
At this point we notice that the form of the DBI action has emerged.  From the form of (\ref{zero_modes}), we see that the $\mathcal{F}$ dependence of the matrix $\cT$ determines the coupling of the boundary state, in the presence of worldvolume flux, to the NS-NS massless modes. Alternatively, this matrix can be obtained directly from the DBI action by expanding it about a (flat) background metric.

With all these ingredients at hand, we are ready to compute the interaction between a D3 and a D7 brane. In the case under consideration, the D3 stretches along $\{x^5...x^8\}$ whereas the D7 is aligned along $\{x^1...x^8\}$.  The background NS two form potential $B_2$ vanishes. We however turn on a non-vanishing (constant) magnetic flux in the $\{x^1...x^4\}$ directions on the D7, which is arranged in a block-diagonal form.
i.e.
\begin{equation}
F_2 = \left(\begin{array}{ccccc}
0 &  F_{12} & 0  &0              &0\\
-F_{12}  & 0 &  0 & 0             &0\\
               0&  0 &0& F_{34} &0\\
               0&  0 &-F_{34} &0&0 \\
               0&  0 & 0             &   0&\mathbf{0_{4\times 4}}
\end{array}
\right).
\end{equation}
In this case we thus have $\mathcal{F} = 2\pi \alpha' F_2$, and so $f_1 = 2\pi\alpha'F_{12}$ and $f_2 = 2\pi\alpha'F_{34}$.
The cylinder diagram is then given by
\begin{equation}
<+,B_{3+1}| \Delta | B_{7+1},+, \mathcal{F}>, \qquad \Delta = \int \frac{d^2z}{|z|^2}  z^{L_0}\bar{z}^{\tilde{L_0}},
\end{equation}
where $\Delta$ is the closed string propagator. It gives directly the coupling of the D3 brane worldvolume scalars to the D7 worldvolume scalars and gauge fields.
The contribution of massless mode (i.e. zero modes) exchange is readily extracted. In fact it is given simply by
\begin{equation}
<+, B_{0(x^5..x^8)} |B_{0(x^1...x^8),} ,+, \mathcal{F}> = T_3 T_{7}\sqrt{\det(1+\mathcal{F})}\left[\textrm{tr}(\mathcal{T}^{3} \mathcal{T}^{7}(\mathcal{F})) + \textrm{tr}(\cM^{3}\cM^{7}(\mathcal{F}))\right].
\end{equation}
The first term gives the contribution of the graviton and dilaton and the second term the R-R potential.
Evaluating explicitly, we have
\begin{equation}
<+,B_{0(x^5..x^8)} |B_{0(x^1...x^8)} ,+, \mathcal{F}> = T_3T_7\left[-4\left(F_{12}^2\sqrt{\frac{1+F_{34}^2}{1+F_{12}^2}} + F_{34}^2 \sqrt{\frac{1+F_{12}^2}{1+F_{34}^2}}\right) - 8F_{12}F_{34}\right].
\end{equation}
As a consistency check, at zero $F$'s, the contribution is trivially zero. This is consistent with the fact that the branes interact only via the exchange of gravitons
and dilatons
and that they are mutually BPS, such that the net force between them vanishes.

When the magnetic flux in $\{x^1,x^2,x^3, x^4\}$ satisfies the anti-self-dual condition, which in the case of ${\mathbb{R}}^4$ is equivalent to
\begin{equation}\label{antiselfdual}
F_{12} + F_{34} = 0,
\end{equation}
the D3 branes become decoupled from the D7 world-volume flux.
The calculation is unaltered if we compactify $\{x^1,x^2,x^3, x^4\}$ on a torus, except that an infinite array of D3s is periodically placed in the covering space.  In particular, the decoupling between the branes under (\ref{antiselfdual}) continues to hold.

\section{Some Geometric Details of the Deformed Conifold}\label{DefconifoldDetails}
\paragraph{}
Here we provide some geometric details about the deformed conifold which were used in deriving the F-term scalar potential in the main text.

The deformed conifold can be defined as a subspace in ${\mathbb{C}}^4$ by the following constraint equation:
\begin{equation}
z_1^2+z_2^2+z_3^2+z_4^2=\epsilon^2\,,\label{Defdefconifold}
\end{equation}
where $z_\alpha\in {\mathbb{C}}\,,~\alpha=1\,,\dots\,,4$ and the deformation parameter $\epsilon\in {\mathbb{R}}$. The radial coordinates can be related to the norm of the embedding coordinate:
\begin{equation}
\sum_{\alpha=1}^4 |z_\alpha|^2=r^3=\epsilon^2\cosh\tau\,.\label{Defradial}
\end{equation}
The K\"ahler metric of the deformed conifold can be derived from the K\"ahler potential:
\begin{equation}
k(\tau)=\frac{\epsilon^{4/3}}{2^{1/3}}\int_\tau\,d\tau' (\sinh(2\tau')-2\tau')^{1/3}.\label{DefconKahlerpotential}
\end{equation}
Using the embedding condition (\ref{Defdefconifold}), we can substitute away $z_4$ as $z_4=\pm\sqrt{\epsilon^2-(z_1^2+z_2^2+z_3^2)}$, and the explicit K\"ahler metric is given by:
\begin{equation}
k_{i\bar{j}}=\partial_i\partial_{\bar{j}}k(\tau)=\bA(\tau) {\mathrm{M}}_{i\bar{j}}+\bB(\tau){\mathrm{N}}_{i\bar{j}}\,,~~~i,\bar{j}=1\,,\dots\,,3\label{Defdefconmetric}
\end{equation}
Here we have applied the chain rule from (\ref{Defradial}), the functions $\bA(\tau)$ and $\bB(\tau)$ are given by:
\begin{equation}
\bA(\tau)=\frac{1}{\epsilon^2\sinh\tau}\frac{\partial k(\tau)}{\partial\tau}\,,~~~\bB(\tau)=\frac{1}{\epsilon^4\sinh^2\tau}\left(\frac{\partial^2 k(\tau)}{\partial\tau^2}-\coth\tau \frac{\partial k(\tau)}{\partial\tau}\right)\,,\label{DefABtau}
\end{equation}
whereas the $3\times 3$ matrices $\mathrm{M}_{i\bar{j}}$ and $\mathrm{N}_{i\bar{j}}$ are given by:
\begin{eqnarray}
\mathrm{M}_{i\bar{j}}&=&\delta_{i\bar{j}}+\frac{z_i\bar{z}_j}{|z_4|^2}\,,\label{DefMij}\\
\mathrm{N}_{i\bar{j}}&=&z_i\bar{z}_j+\bz_i z_j-z_i z_j \frac{\bz_4}{z_4}-\bz_i \bz_j\frac{z_4}{\bz_4}\,.\label{DefNij}
\end{eqnarray}
The inverse K\"ahler metric can be calculated in a tedious but straightforward manner, and the result can be written as:
\begin{equation}
k^{\bar{i}{j}}=\frac{1}{(\bA(\tau)+r^3\tanh^2\tau \bB(\tau))}\left({\mathrm{R}}^{\bar{i}{j}}+\left(\frac{ r^3 \bB(\tau)}{\bA(\tau)}\right){\mathrm{L}}^{\bar{i}{j}}\right)=\frac{r^3}{\frac{\partial^2 k(\tau)}{\partial\tau^2}}\left({\mathrm{R}}^{\bar{i}{j}}+\left(\frac{ r^3 \bB(\tau)}{\bA(\tau)}\right){\mathrm{L}}^{\bar{i}{j}}\right)\label{InvDefConMetric}\,,
\end{equation}
with the $3\times 3$ matrices ${\mathrm{R}}^{\bar{i}{j}}$ and ${\mathrm{L}}^{\bar{i}{j}}$ given by:
\begin{eqnarray}
{\mathrm{R}}^{\bar{i}{j}}&=&\delta^{\bar{i}{j}}-\frac{z_i\bz_j}{r^3}\,,\label{defRij}\\
{\mathrm{L}}^{\bar{i}{j}}&=&\left(1-\frac{\epsilon^4}{r^6}\right)\delta^{\bar{i}{j}}
+\frac{\epsilon^2}{r^3}\left(\frac{z_i z_j+\bz_i\bz_j}{r^3}\right)-\left(\frac{z_i \bz_j+\bz_i z_j}{r^3}\right)\,.\label{defLij}
\end{eqnarray}
From these matrices, the various terms in the scalar potential can be calculated readily. Notice that ${\mathrm{L}}_{i\bar{j}}$ vanishes as $r\to\epsilon^{2/3}$, and the inverse metric (\ref{InvDefConMetric}) reduces to the one derived using the simplified K\"ahler potential near the tip.
While at large $r\gg \epsilon^{2/3}$ and $\epsilon^2(e^\tau/2)\approx r^3$, $k^{\bar{i}j}$ readily reduces to the usual inverse metric for the singular conifold.

One should also note that the complex embedding coordinates can also be expressed in terms of six real coordinates:
$\{\tau\in{\mathbb{R}\,,~\psi\in[0,4\pi]\,,~\theta_{1,2}\in[0,\pi]\,,~\phi_{1,2}\in[0,2\pi]}\}$,
$\Xi=\tau+i\psi$ as
{\small \begin{eqnarray}
\label{FullDefConifoldcoord1}
z_1 &=& \epsilon \left[ \cosh \left(\frac{\Xi}{2}\right) \cos
\left(\frac{\theta_1+\theta_2}{2}\right) \cos\left(\frac{\phi_1+\phi_2}{2}\right) +
i \sinh \left(\frac{\Xi}{2}\right) \cos \left(\frac{\theta_1-\theta_2}{2}\right)
\sin \left(\frac{\phi_1+\phi_2}{2}\right) \right] \, ,\nonumber
\\
\label{FullDefConifoldcoord2}
z_2 &=& \epsilon \left[ -\cosh \left(\frac{\Xi}{2}\right) \cos
\left(\frac{\theta_1+\theta_2}{2}\right) \sin \left(\frac{\phi_1+\phi_2}{2}\right) +
i \sinh \left(\frac{\Xi}{2}\right) \cos \left(\frac{\theta_1-\theta_2}{2}\right)
\cos \left(\frac{\phi_1+\phi_2}{2}\right)\right] \, ,\nonumber
\\
\label{FullDefConifoldcoord3}
z_3 &=& \epsilon \left[ -\cosh \left(\frac{\Xi}{2}\right) \sin
\left(\frac{\theta_1+\theta_2}{2}\right) \cos \left(\frac{\phi_1-\phi_2}{2}\right) +
i \sinh \left(\frac{\Xi}{2}\right) \sin \left(\frac{\theta_1-\theta_2}{2}\right)
\sin \left(\frac{\phi_1-\phi_2}{2}\right)\right] \, ,\nonumber
\\
\label{FullDefConifoldcoord4}
z_4 &=& \epsilon \left[ -\cosh \left(\frac{\Xi}{2}\right) \sin
\left(\frac{\theta_1+\theta_2}{2}\right) \sin \left(\frac{\phi_1-\phi_2}{2}\right) -
i \sinh \left(\frac{\Xi}{2}\right) \sin \left(\frac{\theta_1-\theta_2}{2}\right)
\cos\left(\frac{\phi_1-\phi_2}{2}\right)\right] \, .\nonumber\\
\end{eqnarray}}
$\!\!\!$By substituting (\ref{FullDefConifoldcoord4}) into (\ref{Defdefconmetric}), with appropriate rearrangement, one can obtain the more familiar deformed conifold metric in terms of the usual radial and angular coordinates:
{\small \begin{equation}
ds_6^2 = \frac{1}{2} \epsilon^{4/3} K(\tau) \left\{ \frac{1}{3 [K(\tau)]^3} [d\tau^2
+(g^5)^2] + \cosh^2 \left( \frac{\tau}{2} \right) \left[ (g^3)^2 + (g^4)^2 \right] +
\sinh^2 \left( \frac{\tau}{2} \right) \left[(g^1)^2 + (g^2)^2 \right] \right\} \, .
\label{ExpDeformMetric}
\end{equation}}
$\!\!\!$Here the function $K(\tau)$ is as defined in (\ref{DefKtau}), whereas the one forms are defined as:
\begin{eqnarray}
g_1&=&\frac{1}{\sqrt{2}}(-\sin\theta_1 d\phi_1-(\cos\psi\sin\theta_2 d\phi_2-\sin\psi d\theta_2))\,,
%~~
\nonumber \\
g_2&=&\frac{1}{\sqrt{2}}(d\theta_1-(\sin\psi\sin\theta_2 d\phi_2+\cos\psi d\theta_2))\,,\nonumber\\
g_3&=&\frac{1}{\sqrt{2}}(-\sin\theta_1 d\phi_1+(\cos\psi\sin\theta_2 d\phi_2-\sin\psi d\theta_2))\,,
%~~
\nonumber \\
g_4 &=&\frac{1}{\sqrt{2}}(d\theta_1+(\sin\psi\sin\theta_2 d\phi_2+\cos\psi d\theta_2))\,,\nonumber\\
g_5&=& d\psi+\cos\theta_1 d\phi_1+\cos\theta_2 d\phi_2\,.\label{g1tog5}
\end{eqnarray}
The explicit metric (\ref{ExpDeformMetric}) allows us to derive the expression for the canonical inflaton from the $D3$ brane DBI action, and it would also be interesting to obtain an explicit expression for the inverse deformed conifold metric in terms of the radial and angular coordinates.

\section{Angular Stability Analysis}\label{AngStabAnalysis}
\paragraph{}
In this appendix, we aim to demonstrate that along the trajectory $z_1=-\epsilon\cosh\frac{\tau}{2}$, the angular directions are indeed stabilized by the presence of two stacks of D7 branes, following \cite{Baumann1,CGS1}.
The key ingredient for doing so is the angular mass matrix, and in its diagonal form, the eigenvalues are defined to be:
\begin{eqnarray}
X&=&\mp 2\epsilon\cosh\frac{\tau}{2}\frac{\partial V_F(z_1,\bar{z}_1)}{\partial (z_1+\bar{z}_1)}=\mp 2\epsilon\cosh\frac{\tau}{2}\frac{(z_1\partial_{z_1}-\bz_1\partial_{\bz_1})}{(z_1-\bz_1)}V_F(z_1,\bar{z}_1)\,,\label{DefX}\\
Y&=&-2\left(\epsilon\cosh\frac{\tau}{2}\right)^2\frac{\partial V_F(z_1,\bar{z}_1)}{\partial |z_1|^2}=2\left(\epsilon\cosh\frac{\tau}{2}\right)^2\frac{(\partial_{z_1}-\partial_{\bz_1})}{(z_1-\bz_1)}V_F(z_1,\bar{z}_1)\,,\label{DefY}
\end{eqnarray}
Here we have followed the notations used in \cite{Baumann1,CGS1}, and we have assumed that for the possible field range that inflation can take place, the angular dependences are encoded exclusively in the F-term scalar potential $V_{F}$. This assumption is valid in the region where the Coulombic attraction does not become dominant, as such potential is a function of  the $\rm D3-\overline{\rm D3}$ separation which carries both radial and angular dependences in general \cite{CGS1}.
We shall use the explicit expressions (\ref{DefVKKLT}) and (\ref{DefDeltaV}) for $V_F(z_1,\bz_1)$ and the mass eigenvalues can be readily calculated as:
\begin{eqnarray}
\mp\frac{X}{2\mathcal{C}\cosh\frac{\tau}{2}}&=&\bigg[U(\tau,\sigma) + (2 \beta^2 \sigma_0) P(\tau) + \frac{6}{S_0(\tau)}[(A_0 e^{-a_1\sigma}(1+\alpha_1\cosh(\tau/2))\nonumber \\
&+&B_0 e^{-a_2\sigma}(1+\alpha_2\cosh(\tau/2))]
- 3\frac{|W_0|}{|S_0(\tau)|} \Bigg](\widetilde{(z\Delta)}|S_0(z_1)|^2)\nonumber \\
&+& 6 |S_0(\tau)|^2\widetilde{(z\Delta)}\textrm{Re}\left[\frac{(A_0 e^{-a_1\sigma}(1-z_1/\mu_1)+B_0 e^{-a_2\sigma}(1- z_1/\mu_2))}{S_0(z_1)}\right]\nonumber\\
&+&\frac{\epsilon^{4/3}}{2\beta^2\sigma_0}
\left(|S_1(\tau)|^2(\widetilde{(z\Delta)}T(z_1,\bz_1))+T(\tau)(\widetilde{(z\Delta)}|S_1(z_1)|^2)\right)\nonumber\\
&+&\coth\tau L(\tau)\widetilde{(z\Delta)}
\left(S_0(z_1)\bar{S}_1(\bz_1) \left(\bar{z_1}- \frac{\epsilon^2}{r^3}z_1\right)+\bar{S}_0(\bz_1)S_1(z_1) \left(z_1- \frac{\epsilon^2}{r^3}\bz_1\right)\right)\,,\nonumber\\
\label{MassX}
\end{eqnarray}
\begin{eqnarray}
\frac{Y}{2\mathcal{C}\cosh^2(\frac{\tau}{2})}&=&\bigg[U(\tau,\sigma) + (2 \beta^2 \sigma_0) P(\tau) + \frac{6}{S_0(\tau)}[(A_0 e^{-a_1\sigma}(1+\alpha_1 \cosh(\tau/2))\nonumber \\
&+& B_0 e^{-a_2\sigma}(1+\alpha_2 \cosh(\tau/2)))]
- 3\frac{|W_0|}{|S_0(\tau)|} \Bigg](\widetilde{\Delta}|S_0(z_1)|^2)\nonumber \\
&+& 6 |S_0(\tau)|^2\widetilde{\Delta}\textrm{Re}[\frac{(A_0 e^{-a_1\sigma}(1-z_1/\mu_1)+B_0 e^{-a_2\sigma}(1- z_1/\mu_2))}{S_0(z_1)}]\nonumber\\
&+&\frac{\epsilon^{4/3}}{2\beta^2 \sigma_0}
\left(|S_1(\tau)|^2(\widetilde{\Delta}T(z_1,\bz_1))+T(\tau)(\widetilde{\Delta}|S_1(z_1)|^2)\right)\nonumber\\
&+&\coth\tau L(\tau)\widetilde{\Delta}
\left(S_0(z_1)\bar{S}_1(\bz_1) \left(\bar{z_1}- \frac{\epsilon^2}{r^3}z_1\right)+\bar{S}_0(\bz_1)S_1(z_1) \left(z_1- \frac{\epsilon^2}{r^3}\bz_1\right)\right)\,,\nonumber\\
\label{MassY}
\end{eqnarray}
where $\mathcal{C} = \frac{\epsilon \kappa^2}{3U(\tau,\sigma)^2}$.
Here we have defined the differential operators:
\begin{equation}
\widetilde{(z\Delta)}=\frac{(z_1\partial_{z_1}-\bz_1\partial_{\bz_1})}{(z_1-\bz_1)}\,,~~
\widetilde{\Delta}=\frac{(\partial_{z_1}-\partial_{\bz_1})}{(z_1-\bz_1)}\,.
\end{equation}
which act on expressions involving $\{z_1,\bar{z}_1\}$. It is understood that  (\ref{MassX}) and (\ref{MassY}) are evaluated along the trajectory $z_1=-\epsilon\cosh\frac{\tau}{2}$, therefore we have substituted the terms in (\ref{MassX}) and (\ref{MassY}) that do not involve the derivatives; whereas the various terms involving derivatives for $A_0\,,B_0\in {\mathbb{R}}$ can be obtained from a tedious but straightforward computer-aided calculation. Here we shall not list out the actual expressions as they are long and not very illuminating. However, for demonstration purpose, we shall consider the small and large radius limits, where the analysis simplifies. We also numerically checked that in the intermediate region, angular stability can be achieved.
%%%%%%%%%%%%%%%%%%%%%%%%%%%%%%%%%%%%%%%%%%%%%%%%%%%%%%%%%%%%%%%%%%%%%%%%%%%%%%%
\subsection{Near tip limit}
\paragraph{}
Let us first consider in the near tip limit $\tau\to\ 0$, where all the mass eigenvalues reduce smoothly to: \cite{CGS1}
\begin{equation}
M = X+Y \,,
\end{equation}
evaluated along the trajectory $z_1 = - \epsilon \cosh(\tau/2)$.
This is given simply by
\begin{eqnarray}
M&\sim&  \mathcal{C}  \Bigg[U(0,\sigma) + (2 \beta^2 \sigma_0) P(0) + \bigg(6\left[\frac{A_0 e^{-a_1\sigma}g_1(0)+B_0 e^{-a_2\sigma}g_2(0)}{S_0(0)}\right]\nonumber \\
&-& 3\frac{|W_0|}{|S_0(0)|} \bigg)\Bigg] [-2 (\widetilde{z\Delta}+\widetilde{\Delta})|S_0(0)|^2] \nonumber \\
&+& 6|S_0(0)|^2(\widetilde{z\Delta}+\widetilde{\Delta})\left[\frac{A_0 e^{-a_1\sigma}g_1(0)+B_0 e^{-a_2\sigma}g_2(0)}{S_0(0)}\right]
\,.
\end{eqnarray}
The terms in the last line are relatively small, especially when $n_1 \sim n_2$. The mass is dominated by terms in the first two lines. The expression in the large square bracket is generally negative. We therefore need $[-2 (\widetilde{z\Delta}+\widetilde{\Delta})|S_0(0)|^2]$ to be negative, which is easily satisfied.

%%%%%%%%%%%%%%%%%%%%%%%%%%%%%%%%%%%%%%%%%%%%%%%%%%%
\subsection{Large radius limit}
\paragraph{}
In the large radius limit, the terms that dominate in the small radius limit remain dominant for positive $A_0,B_0,\alpha_1$ and $\alpha_2$. Like the near tip limit, the stability depends crucially on the sign of the terms in the square bracket in (\ref{MassX}), which is generally negative for small $\tau$ and becomes positive for larger radii. Given that $\widetilde{z \Delta}|S_0(z_1)|^2$ and $\widetilde{\Delta}|S_0(z_1)|^2$ are generally negative, this means stability is ensured in the small radius limit, but becomes unstable for sufficiently large radius. The presence of the extra terms serve to extend the region of stability. The precise positions where the eigenvalues hit zero depends on the choice of parameters. In general for smaller $\beta$, the extra terms involving $T(z_1,\bar{z}_1)$ and its derivatives are enhanced, such that the masses remain positive up to larger $\tau$. For $n_1 \ne n_2$, i.e. Case C, the hierarchy between terms is not as obvious and requires an explicit check. We have checked the masses for all the cases considered, and found angular stability within the region where inflation takes place. Notice that to achieve such stability, it is crucial to take the adiabatic approximation (\ref{Defsigmastar}) and use the numerical solution of $\sigma_\star(\tau)$.
%%%%%%%%%%%%%%%%%%%%%%%%%%%%%%%%%%%%%%%%%%%%%%%%%%%%%%%%%%%%%%%%%%%%%%%%%%%%%%%
\begin{figure}[htp]
\centering
\includegraphics[
width=0.4\textwidth]{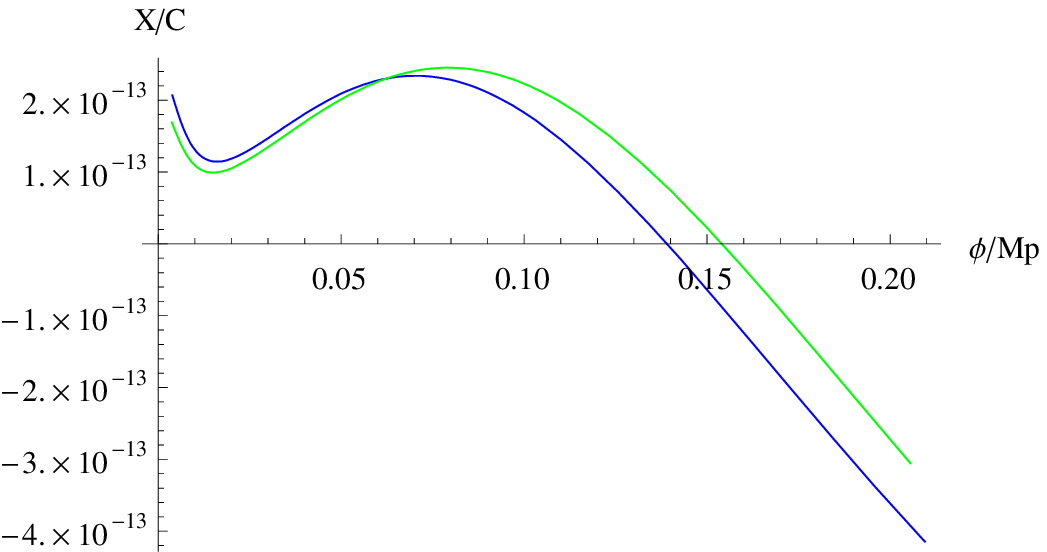}
\includegraphics[
width=0.4\textwidth]{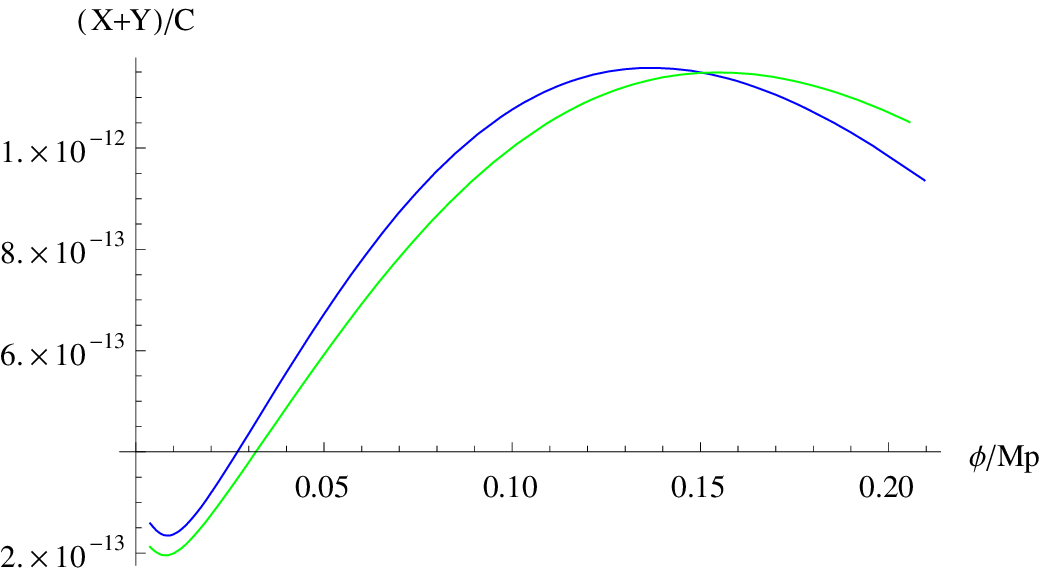}
\caption{Mass eigenvalues for Case A (blue) and Case B (green)}
\end{figure}
%%%%%%%%%%%%%%%%%%%%%%%%%%%%%%%%%%%%%%%%%%%%%%%%%%%%%%%%%%%%%%%%%%%%%%%%%%%%%%%
\begin{figure}[htp]
\centering
\includegraphics[
width=0.4\textwidth]{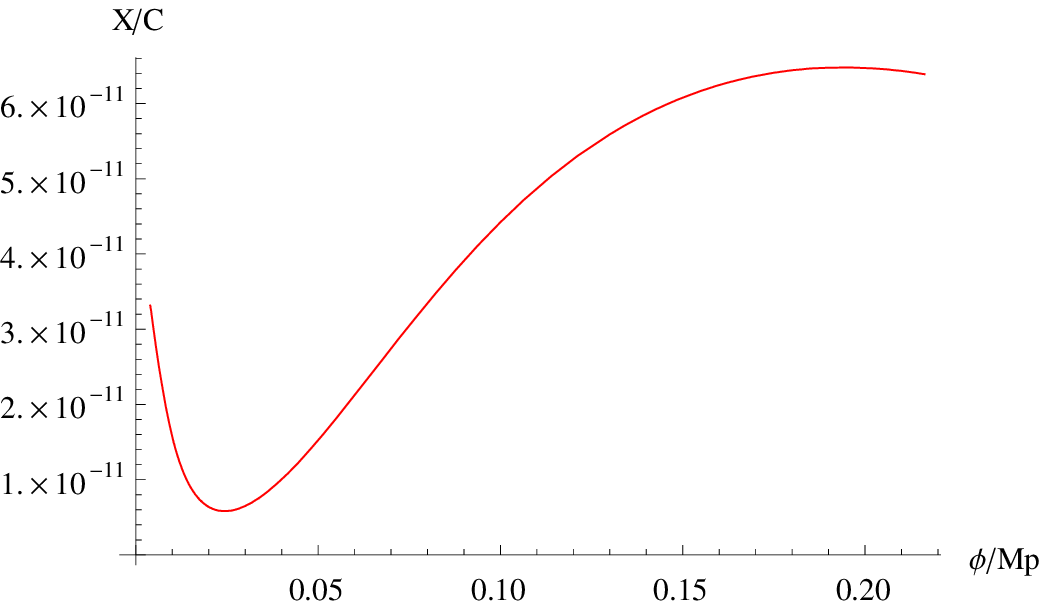}
\includegraphics[
width=0.4\textwidth]{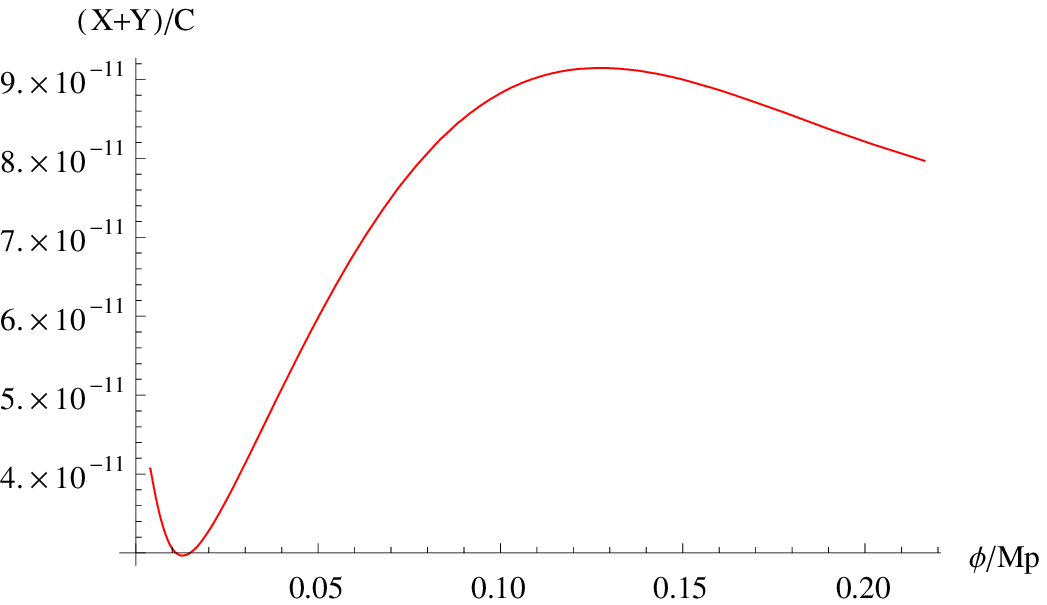}
\caption{Mass eigenvalues for Case C}
\end{figure}


\newpage


\begin{thebibliography}{99}

\bibitem{inflation}
  A.~H.~Guth,
  %``The Inflationary Universe: A Possible Solution To The Horizon And Flatness
  %Problems,''
  Phys.\ Rev.\  D {\bf 23}, 347 (1981).
  %%CITATION = PHRVA,D23,347;%%
  A.~D.~Linde,
  %``A New Inflationary Universe Scenario: A Possible Solution Of The Horizon,
  %Flatness, Homogeneity, Isotropy And Primordial Monopole Problems,''
  Phys.\ Lett.\  B {\bf 108}, 389 (1982)~;
  %%CITATION = PHLTA,B108,389;%%
  A.~Albrecht and P.~J.~Steinhardt,
  %``Cosmology For Grand Unified Theories With Radiatively Induced Symmetry
  %Breaking,''
  Phys.\ Rev.\ Lett.\  {\bf 48}, 1220 (1982).
  %%CITATION = PRLTA,48,1220;%%




\bibitem{KKLMMT}
  S.~Kachru, R.~Kallosh, A.~Linde, J.~M.~Maldacena, L.~P.~McAllister and S.~P.~Trivedi,
  %``Towards inflation in string theory,''
  JCAP {\bf 0310} (2003) 013
  [arXiv:hep-th/0308055].
  %%CITATION = JCAPA,0310,013;%%


\bibitem{McGill1}
  C.~P.~Burgess, J.~M.~Cline, K.~Dasgupta and H.~Firouzjahi,
  %``Uplifting and inflation with D3 branes,''
  JHEP {\bf 0703}, 027 (2007)
  [arXiv:hep-th/0610320].
  %%CITATION = JHEPA,0703,027;%%




%\cite{Baumann:2007np}
\bibitem{Baumann:2007np}
  D.~Baumann, A.~Dymarsky, I.~R.~Klebanov, L.~McAllister and P.~J.~Steinhardt,
  %``A Delicate Universe,''
  Phys.\ Rev.\ Lett.\  {\bf 99}, 141601 (2007)
  [arXiv:0705.3837 [hep-th]].
  %%CITATION = PRLTA,99,141601;%%

 %\cite{Krause:2007jk}
\bibitem{Krause:2007jk}
  A.~Krause and E.~Pajer,
  %``Chasing Brane Inflation in String-Theory,''
  JCAP {\bf 0807}, 023 (2008)
  [arXiv:0705.4682 [hep-th]].
  %%CITATION = JCAPA,0807,023;%%

\bibitem{Baumann1}
  D.~Baumann, A.~Dymarsky, I.~R.~Klebanov and L.~McAllister,
  %``Towards an Explicit Model of D-brane Inflation,''
  JCAP {\bf 0801} (2008) 024
  [arXiv:0706.0360 [hep-th]].
  %%CITATION = JCAPA,0801,024;%%
%\cite{Klebanov:2000hb}

\bibitem{HoloBulkeffects}
  D.~Baumann, A.~Dymarsky, S.~Kachru, I.~R.~Klebanov and L.~McAllister,
  %``Holographic Systematics of D-brane Inflation,''
  arXiv:0808.2811 [hep-th].
  %%CITATION = ARXIV:0808.2811;%%



%\cite{Baumann:2006cd}
\bibitem{Baumann:2006cd}
  D.~Baumann and L.~McAllister,
  %``A Microscopic Limit on Gravitational Waves from D-brane Inflation,''
  Phys.\ Rev.\  D {\bf 75}, 123508 (2007)
  [arXiv:hep-th/0610285].
  %%CITATION = PHRVA,D75,123508;%%

  %\cite{Kobayashi:2007hm}
\bibitem{Kobayashi:2007hm}
  T.~Kobayashi, S.~Mukohyama and S.~Kinoshita,
  %``Constraints on Wrapped DBI Inflation in a Warped Throat,''
  JCAP {\bf 0801}, 028 (2008)
  [arXiv:0708.4285 [hep-th]].
  %%CITATION = JCAPA,0801,028;%%

  %\cite{Becker:2007ui}
\bibitem{Becker:2007ui}
  M.~Becker, L.~Leblond and S.~E.~Shandera,
  %``Inflation from Wrapped Branes,''
  Phys.\ Rev.\  D {\bf 76}, 123516 (2007)
  [arXiv:0709.1170 [hep-th]].
  %%CITATION = PHRVA,D76,123516;%%


%\cite{Kallosh:2007cc}
\bibitem{Kallosh:2007cc}
  R.~Kallosh, N.~Sivanandam and M.~Soroush,
  %``Axion Inflation and Gravity Waves in String Theory,''
  Phys.\ Rev.\  D {\bf 77}, 043501 (2008)
  [arXiv:0710.3429 [hep-th]].
  %%CITATION = PHRVA,D77,043501;%%

  %\cite{Grimm:2007hs}
\bibitem{Grimm:2007hs}
  T.~W.~Grimm,
  %``Axion Inflation in Type II String Theory,''
  Phys.\ Rev.\  D {\bf 77}, 126007 (2008)
  [arXiv:0710.3883 [hep-th]].
  %%CITATION = PHRVA,D77,126007;%%

%\cite{Silverstein:2008sg}
\bibitem{Silverstein:2008sg}
  E.~Silverstein and A.~Westphal,
  %``Monodromy in the CMB: Gravity Waves and String Inflation,''
  Phys.\ Rev.\  D {\bf 78}, 106003 (2008)
  [arXiv:0803.3085 [hep-th]].
  %%CITATION = PHRVA,D78,106003;%%


  %\cite{McAllister:2008hb}
\bibitem{McAllister:2008hb}
  L.~McAllister, E.~Silverstein and A.~Westphal,
  %``Gravity Waves and Linear Inflation from Axion Monodromy,''
  arXiv:0808.0706 [hep-th].
  %%CITATION = ARXIV:0808.0706;%%

%\cite{Cicoli:2008gp}
\bibitem{Cicoli:2008gp}
  M.~Cicoli, C.~P.~Burgess and F.~Quevedo,
  %``Fibre Inflation: Observable Gravity Waves from IIB String
  %Compactifications,''
  arXiv:0808.0691 [hep-th].
  %%CITATION = ARXIV:0808.0691;%%

  %\cite{Kecskemeti:2006cg}
\bibitem{Kecskemeti:2006cg}
  S.~Kecskemeti, J.~Maiden, G.~Shiu and B.~Underwood,
  %``DBI inflation in the tip region of a warped throat,''
  JHEP {\bf 0609}, 076 (2006)
  [arXiv:hep-th/0605189].
  %%CITATION = JHEPA,0609,076;%%

%\cite{Shiu:2006kj}
\bibitem{Shiu:2006kj}
  G.~Shiu and B.~Underwood,
  %``Observing the Geometry of Warped Compactification via Cosmic Inflation,''
  Phys.\ Rev.\ Lett.\  {\bf 98}, 051301 (2007)
  [arXiv:hep-th/0610151].
  %%CITATION = PRLTA,98,051301;%%
%\cite{DeWolfe:2007hd}
\bibitem{DeWolfe:2007hd}
  O.~DeWolfe, L.~McAllister, G.~Shiu and B.~Underwood,
  %``D3-brane Vacua in Stabilized Compactifications,''
  JHEP {\bf 0709}, 121 (2007)
  [arXiv:hep-th/0703088].
  %%CITATION = JHEPA,0709,121;%%

%\cite{Easson:2007dh}
\bibitem{Easson:2007dh}
  D.~A.~Easson, R.~Gregory, D.~F.~Mota, G.~Tasinato and I.~Zavala,
  %``Spinflation,''
  JCAP {\bf 0802}, 010 (2008)
  [arXiv:0709.2666 [hep-th]].
  %%CITATION = JCAPA,0802,010;%%



%\cite{Huang:2007hh}
\bibitem{Huang:2007hh}
  M.~x.~Huang, G.~Shiu and B.~Underwood,
  %``Multifield DBI Inflation and Non-Gaussianities,''
  Phys.\ Rev.\  D {\bf 77}, 023511 (2008)
  [arXiv:0709.3299 [hep-th]].
  %%CITATION = PHRVA,D77,023511;%%

  \bibitem{CGS1}
  H.~Y.~Chen, J.~O.~Gong and G.~Shiu,
  %``Systematics of multi-field effects at the end of warped brane inflation,''
  arXiv:0807.1927 [hep-th].
  %%CITATION = ARXIV:0807.1927;%%

\bibitem{KLbound}
  R.~Kallosh and A.~Linde,
  %``Landscape, the scale of SUSY breaking, and inflation,''
  JHEP {\bf 0412}, 004 (2004)
  [arXiv:hep-th/0411011].
  %%CITATION = JHEPA,0412,004;%%

\bibitem{KKLT}
  S.~Kachru, R.~Kallosh, A.~Linde and S.~P.~Trivedi,
  %``De Sitter vacua in string theory,''
  Phys.\ Rev.\  D {\bf 68}, 046005 (2003)
  [arXiv:hep-th/0301240].
  %%CITATION = PHRVA,D68,046005;%%

%\cite{Conlon:2008cj}
\bibitem{CKLQ}
  J.~P.~Conlon, R.~Kallosh, A.~Linde and F.~Quevedo,
  %``Volume Modulus Inflation and the Gravitino Mass Problem,''
  JCAP {\bf 0809}, 011 (2008)
  [arXiv:0806.0809 [hep-th]].
  %%CITATION = JCAPA,0809,011;%%


%\cite{Kallosh:2007wm}
\bibitem{KL2}
  R.~Kallosh and A.~Linde,
  %``Testing String Theory with CMB,''
  JCAP {\bf 0704}, 017 (2007)
  [arXiv:0704.0647 [hep-th]].
  %%CITATION = JCAPA,0704,017;%%



\bibitem{AccidentalInflation}
  A.~Linde and A.~Westphal,
  %``Accidental Inflation in String Theory,''
  JCAP {\bf 0803}, 005 (2008)
  [arXiv:0712.1610 [hep-th]].
  %%CITATION = JCAPA,0803,005;%%

  \bibitem{BO1}
  M.~Badziak and M.~Olechowski,
  %``Volume modulus inflation and a low scale of SUSY breaking,''
  JCAP {\bf 0807}, 021 (2008)
  [arXiv:0802.1014 [hep-th]].
  %%CITATION = JCAPA,0807,021;%%

%\cite{Badziak:2008gv}
\bibitem{BO2}
  M.~Badziak and M.~Olechowski,
  %``Volume modulus inflection point inflation and the gravitino mass problem,''
  arXiv:0810.4251 [hep-th].
  %%CITATION = ARXIV:0810.4251;%%


%\cite{Abe:2008xu}
\bibitem{Abe:2008xu}
  H.~Abe, T.~Higaki, T.~Kobayashi and O.~Seto,
  %``Non-perturbative moduli superpotential with positive exponents,''
  Phys.\ Rev.\  D {\bf 78}, 025007 (2008)
  [arXiv:0804.3229 [hep-th]].
  %%CITATION = PHRVA,D78,025007;%%

%\cite{Dvali:1998pa}
\bibitem{Dvali:1998pa}
  G.~R.~Dvali and S.~H.~H.~Tye,
  %``Brane inflation,''
  Phys.\ Lett.\  B {\bf 450}, 72 (1999)
  [arXiv:hep-ph/9812483].
  %%CITATION = PHLTA,B450,72;%%

\bibitem{RacetrackUnpublished}
 L.~Dixon, V.~Kaplunovsky and M.~Peskin, unpublished; L.J.~Dixon, SLAC-PUB-5229, 1990.

\bibitem{Krasnikov:1987jj}
  N.~V.~Krasnikov,
  %``On Supersymmetry Breaking in Superstring Theories,''
  Phys.\ Lett.\  B {\bf 193} (1987) 37.
  %%CITATION = PHLTA,B193,37;%%

 %\cite{Casas:1990qi}
\bibitem{Casas:1990qi}
  J.~A.~Casas, Z.~Lalak, C.~Munoz and G.~G.~Ross,
  %``HIERARCHICAL SUPERSYMMETRY BREAKING AND DYNAMICAL DETERMINATION OF
  %COMPACTIFICATION PARAMETERS BY NONPERTURBATIVE EFFECTS,''
  Nucl.\ Phys.\  B {\bf 347}, 243 (1990).
  %%CITATION = NUPHA,B347,243;%%

 %\cite{Taylor:1990wr}
\bibitem{Taylor:1990wr}
  T.~R.~Taylor,
  %``Dilaton, gaugino condensation and supersymmetry breaking,''
  Phys.\ Lett.\  B {\bf 252}, 59 (1990).
  %%CITATION = PHLTA,B252,59;%%

 \bibitem{RaceTrack1}
  J.~J.~Blanco-Pillado {\it et al.},
  %``Racetrack inflation,''
  JHEP {\bf 0411} (2004) 063
  [arXiv:hep-th/0406230].
  %%CITATION = JHEPA,0411,063;%%


 \bibitem{MarchesanoShiu}
% \cite{Marchesano:2004xz}
%\bibitem{Marchesano:2004xz}
  F.~Marchesano and G.~Shiu,
  %``Building MSSM flux vacua,''
  JHEP {\bf 0411}, 041 (2004)
  [arXiv:hep-th/0409132];
  %%CITATION = JHEPA,0411,041;%%
%\cite{Marchesano:2004yq}
%\bibitem{Marchesano:2004yq}
  F.~Marchesano and G.~Shiu,
  %``MSSM vacua from flux compactifications,''
  Phys.\ Rev.\  D {\bf 71}, 011701 (2005)
  [arXiv:hep-th/0408059].
  %%CITATION = PHRVA,D71,011701;%%

\bibitem{Kuperstein}
S.~Kuperstein,
%``Meson spectroscopy from holomorphic probes on the warped deformed %conifold,''
JHEP {\bf 0503} (2005) 014
[arXiv:hep-th/0411097].
%%CITATION = JHEPA,0503,014;%%



 \bibitem{fluxreviews}
  For recent reviews, see, e.g.,
  %\cite{Douglas:2006es}
%\bibitem{Douglas:2006es}
  M.~R.~Douglas and S.~Kachru,
  %``Flux compactification,''
  Rev.\ Mod.\ Phys.\  {\bf 79}, 733 (2007)
  [arXiv:hep-th/0610102];
  %%CITATION = RMPHA,79,733;%%
    %\cite{Blumenhagen:2006ci}
%\bibitem{Blumenhagen:2006ci}
  R.~Blumenhagen, B.~Kors, D.~Lust and S.~Stieberger,
  %``Four-dimensional String Compactifications with D-Branes, Orientifolds   and
  %Fluxes,''
  Phys.\ Rept.\  {\bf 445}, 1 (2007)
  [arXiv:hep-th/0610327].
  %%CITATION = PRPLC,445,1;%%


\bibitem{verlinde99}
  H.~L.~Verlinde,
%  {\em ``Holography and compactification,''}
  Nucl.\ Phys.\  B {\bf 580}, 264 (2000)
  [arXiv:hep-th/9906182].
  %%CITATION = NUPHA,B580,264;%%


\bibitem{Dasgupta:1999ss}
  K.~Dasgupta, G.~Rajesh and S.~Sethi,
 % {\em ``M theory, orientifolds and G-flux,''}
  JHEP {\bf 9908}, 023 (1999)
  [arXiv:hep-th/9908088].
  %%CITATION = JHEPA,9908,023;%%

\bibitem{Greene:2000gh}
  B.~R.~Greene, K.~Schalm and G.~Shiu,
%  {\em ``Warped compactifications in M and F theory,''}
  Nucl.\ Phys.\  B {\bf 584}, 480 (2000)
  [arXiv:hep-th/0004103].
  %%CITATION = NUPHA,B584,480;%%

\bibitem{Becker:2000rz}
  K.~Becker and M.~Becker,
 % {\em ``Compactifying M-theory to four dimensions,''}
  JHEP {\bf 0011}, 029 (2000)
  [arXiv:hep-th/0010282],
  %%CITATION = JHEPA,0011,029;%%
  %{\em ``M-Theory on Eight-Manifolds,''}
  Nucl.\ Phys.\  B {\bf 477}, 155 (1996)
  [arXiv:hep-th/9605053].
  %%CITATION = NUPHA,B477,155;%%


%\cite{Giddings:2001yu}
\bibitem{GKP}
  S.~B.~Giddings, S.~Kachru and J.~Polchinski,
  %``Hierarchies from fluxes in string compactifications,''
  Phys.\ Rev.\  D {\bf 66}, 106006 (2002)
  [arXiv:hep-th/0105097].
  %%CITATION = PHRVA,D66,106006;%%




\bibitem{DeWolfeGiddings}
  O.~DeWolfe and S.~B.~Giddings,
  %``Scales and hierarchies in warped compactifications and brane worlds,''
  Phys.\ Rev.\  D {\bf 67}, 066008 (2003)
  [arXiv:hep-th/0208123].
  %%CITATION = PHRVA,D67,066008;%%

%\cite{Giddings:2005ff}
\bibitem{Giddings:2005ff}
  S.~B.~Giddings and A.~Maharana,
  %``Dynamics of warped compactifications and the shape of the warped
  %landscape,''
  Phys.\ Rev.\  D {\bf 73}, 126003 (2006)
  [arXiv:hep-th/0507158].
  %%CITATION = PHRVA,D73,126003;%%

%\cite{Frey:2006wv}
\bibitem{Frey:2006wv}
  A.~R.~Frey and A.~Maharana,
  %``Warped spectroscopy: Localization of frozen bulk modes,''
  JHEP {\bf 0608}, 021 (2006)
  [arXiv:hep-th/0603233].
  %%CITATION = JHEPA,0608,021;%%

%\cite{Burgess:2006mn}
\bibitem{Burgess:2006mn}
  C.~P.~Burgess, P.~G.~Camara, S.~P.~de Alwis, S.~B.~Giddings, A.~Maharana, F.~Quevedo and K.~Suruliz,
  %``Warped supersymmetry breaking,''
  JHEP {\bf 0804}, 053 (2008)
  [arXiv:hep-th/0610255].
  %%CITATION = JHEPA,0804,053;%%


%\cite{Shiu:2008ry}
\bibitem{Shiu:2008ry}
  G.~Shiu, G.~Torroba, B.~Underwood and M.~R.~Douglas,
  %``Dynamics of Warped Flux Compactifications,''
  JHEP {\bf 0806}, 024 (2008)
  [arXiv:0803.3068 [hep-th]].
  %%CITATION = JHEPA,0806,024;%%

  %\cite{Douglas:2008jx}
\bibitem{Douglas:2008jx}
  M.~R.~Douglas and G.~Torroba,
  %``Kinetic terms in warped compactifications,''
  arXiv:0805.3700 [hep-th].
  %%CITATION = ARXIV:0805.3700;%%

%\cite{Frey:2008xw}
\bibitem{Frey:2008xw}
  A.~R.~Frey, G.~Torroba, B.~Underwood and M.~R.~Douglas,
  %``The Universal Kaehler Modulus in Warped Compactifications,''
  arXiv:0810.5768 [hep-th].
  %%CITATION = ARXIV:0810.5768;%%

%\cite{Marchesano:2008rg}
\bibitem{Marchesano:2008rg}
  F.~Marchesano, P.~McGuirk and G.~Shiu,
  %``Open String Wavefunctions in Warped Compactifications,''
  arXiv:0812.2247 [hep-th].
  %%CITATION = ARXIV:0812.2247;%%

 \bibitem{GKW}
S.~Gukov, C.~Vafa and E.~Witten,
  %``CFT's from Calabi-Yau four-folds,''
  Nucl.\ Phys.\  B {\bf 584} (2000) 69
  [Erratum-ibid.\  B {\bf 608} (2001) 477]
  [arXiv:hep-th/9906070].
  %%CITATION = NUPHA,B584,69;%%


\bibitem{Gaugino1}
  D.~Baumann, A.~Dymarsky, I.~R.~Klebanov, J.~M.~Maldacena, L.~P.~McAllister and A.~Murugan,
  %``On D3-brane potentials in compactifications with fluxes and wrapped
  %D-branes,''
  JHEP {\bf 0611}, 031 (2006)
  [arXiv:hep-th/0607050].
  %%CITATION = JHEPA,0611,031;%%

\bibitem{Ganor}
  O.~J.~Ganor,
  %``A note on zeroes of superpotentials in F-theory,''
  Nucl.\ Phys.\  B {\bf 499} (1997) 55
  [arXiv:hep-th/9612077].
  %%CITATION = NUPHA,B499,55;%%


\bibitem{Berg}
  M.~Berg, M.~Haack and B.~Kors,
  %``Loop corrections to volume moduli and inflation in string theory,''
  Phys.\ Rev.\  D {\bf 71} (2005) 026005
  [arXiv:hep-th/0404087].
  %%CITATION = PHRVA,D71,026005;%%


 \bibitem{COSD7}
  H.~Y.~Chen, P.~Ouyang and G.~Shiu,
  %``On Supersymmetric D7-branes in the Warped Deformed Conifold,''
  arXiv:0807.2428 [hep-th].
  %%CITATION = ARXIV:0807.2428;%%

\bibitem{stringcosmologyreviews}
For recent reviews on the subject of string cosmology, see, e.g.,
  F.~Quevedo,
  %``Lectures on string/brane cosmology,''
  Class.\ Quant.\ Grav.\  {\bf 19}, 5721 (2002)
  [arXiv:hep-th/0210292];
  %%CITATION = CQGRD,19,5721;%%
  A.~Linde,
%  {\em ``Inflation and string cosmology,''}
  eConf {\bf C040802}, L024 (2004)
  [J.\ Phys.\ Conf.\ Ser.\  {\bf 24}, 151 (2005\ PTPSA,163,295-322.2006)]
  [arXiv:hep-th0503195];
  %%CITATION = HEP-TH 0503195;%%
  S.~H.~Henry Tye,
  %{\em ``Brane inflation: String theory viewed from the cosmos,''}
  arXiv:hep-th/0610221;
  %%CITATION = HEP-TH 0610221;%%
  J.~M.~Cline,
  %{\em ``String cosmology,''}
  arXiv:hep-th/0612129;
  %%CITATION = HEP-TH 0612129;%%
  R.~Kallosh,
  %{\em ``On Inflation in String Theory,''}
  arXiv:hep-th/0702059;
  C.~P.~Burgess,
  %{\em ``Lectures on Cosmic Inflation and its Potential Stringy Realizations,''}
  PoS {\bf P2GC}, 008 (2006)
  [Class.\ Quant.\ Grav.\  {\bf 24}, S795 (2007)]
  [arXiv:0708.2865 [hep-th]];
  %%CITATION = CQGRD,24,S795;%%
  L.~McAllister and E.~Silverstein,
  %{\em ``String Cosmology: A Review,''}
  arXiv:0710.2951 [hep-th].
  %%CITATION = ARXIV:0710.2951;%%

%\cite{Gomis:2005wc}
\bibitem{Gomis:2005wc}
  J.~Gomis, F.~Marchesano and D.~Mateos,
  %``An open string landscape,''
  JHEP {\bf 0511}, 021 (2005)
  [arXiv:hep-th/0506179].
  %%CITATION = JHEPA,0511,021;%%

%\cite{Blumenhagen:2005mu}
\bibitem{Blumenhagen:2005mu}
For a review, see, e.g.,
  R.~Blumenhagen, M.~Cvetic, P.~Langacker and G.~Shiu,
  %``Toward realistic intersecting D-brane models,''
  Ann.\ Rev.\ Nucl.\ Part.\ Sci.\  {\bf 55}, 71 (2005)
  [arXiv:hep-th/0502005].
    %%CITATION = ARNUA,55,71;%%



\bibitem{ConifoldNotes}
  P.~Candelas and X.~C.~de la Ossa,
  %``Comments on Conifolds,''
  Nucl.\ Phys.\  B {\bf 342} (1990) 246.
  %%CITATION = NUPHA,B342,246;%%

\bibitem{BKQ}
  C.~P.~Burgess, R.~Kallosh and F.~Quevedo,
  %``de Sitter string vacua from supersymmetric D-terms,''
  JHEP {\bf 0310}, 056 (2003)
  [arXiv:hep-th/0309187].
  %%CITATION = JHEPA,0310,056;%%


\bibitem{KLbound2}
  J.~J.~Blanco-Pillado, R.~Kallosh and A.~Linde,
  %``Supersymmetry and stability of flux vacua,''
  JHEP {\bf 0605} (2006) 053
  [arXiv:hep-th/0511042].
  %%CITATION = JHEPA,0605,053;%%



\bibitem{Overshoot1}
  N.~Itzhaki and E.~D.~Kovetz,
  %``Inflection Point Inflation and Time Dependent Potentials in String
  %Theory,''
  JHEP {\bf 0710}, 054 (2007)
  [arXiv:0708.2798 [hep-th]].
  %%CITATION = JHEPA,0710,054;%%

%\cite{Underwood:2008dh}
\bibitem{Underwood:2008dh}
  B.~Underwood,
  %``Brane Inflation is Attractive,''
  Phys.\ Rev.\  D {\bf 78}, 023509 (2008)
  [arXiv:0802.2117 [hep-th]].
  %%CITATION = PHRVA,D78,023509;%%

%\cite{Hoi:2008gc}
\bibitem{Hoi:2008gc}
  L.~Hoi and J.~M.~Cline,
  %``How Delicate is Brane-Antibrane Inflation?,''
  arXiv:0810.1303 [hep-th].
  %%CITATION = ARXIV:0810.1303;%%

%\cite{Chen:2008ai}
\bibitem{CG1}
  H.~Y.~Chen and J.~O.~Gong,
  %``Towards a warped inflationary brane scanning,''
  arXiv:0812.4649 [hep-th].
  %%CITATION = ARXIV:0812.4649;%%

%\cite{Chen:2006nt}
\bibitem{Chen:2006nt}
  X.~Chen, M.~x.~Huang, S.~Kachru and G.~Shiu,
  %``Observational signatures and non-Gaussianities of general single field
  %inflation,''
  JCAP {\bf 0701}, 002 (2007)
  [arXiv:hep-th/0605045].
  %%CITATION = JCAPA,0701,002;%%
  
%\cite{Langlois:2008qf}
\bibitem{Langlois:2008qf}
  D.~Langlois, S.~Renaux-Petel, D.~A.~Steer and T.~Tanaka,
  %``Primordial perturbations and non-Gaussianities in DBI and general
  %multi-field inflation,''
  Phys.\ Rev.\  D {\bf 78}, 063523 (2008)
  [arXiv:0806.0336 [hep-th]].
  %%CITATION = PHRVA,D78,063523;%%

%\cite{Arroja:2008yy}
\bibitem{Arroja:2008yy}
  F.~Arroja, S.~Mizuno and K.~Koyama,
  %``Non-gaussianity from the bispectrum in general multiple field inflation,''
  JCAP {\bf 0808}, 015 (2008)
  [arXiv:0806.0619 [astro-ph]].
  %%CITATION = JCAPA,0808,015;%%
  
  %\cite{Gao:2008dt}
\bibitem{Gao:2008dt}
  X.~Gao,
  %``Primordial Non-Gaussianities of General Multiple Field Inflation,''
  JCAP {\bf 0806}, 029 (2008)
  [arXiv:0804.1055 [astro-ph]].
  %%CITATION = JCAPA,0806,029;%%
  
  %\cite{Naruko:2008sq}
\bibitem{Naruko:2008sq}
  A.~Naruko and M.~Sasaki,
  %``Large non-Gaussianity from multi-brid inflation,''
  arXiv:0807.0180 [astro-ph].
  %%CITATION = ARXIV:0807.0180;%%


\bibitem{Diaconescu:2005pc}
  D.~E.~Diaconescu, B.~Florea, S.~Kachru and P.~Svrcek,
  ``Gauge - mediated supersymmetry breaking in string compactifications,''
  JHEP {\bf 0602}, 020 (2006)
  [arXiv:hep-th/0512170].
  %%CITATION = JHEPA,0602,020;%%


\bibitem{KS}
  I.~R.~Klebanov and M.~J.~Strassler,
  %``Supergravity and a confining gauge theory: Duality cascades and
  %chiSB-resolution of naked singularities,''
  JHEP {\bf 0008} (2000) 052
  [arXiv:hep-th/0007191].
  %%CITATION = JHEPA,0008,052;%%

\bibitem{Gutperle}
 M.~B.~Green and M.~Gutperle,
 ``Light-cone supersymmetry and D-branes,''
  Nucl.\ Phys.\  B {\bf 476} (1996) 484
  [arXiv:hep-th/9604091].

\bibitem{GSW}
M.B. Green, J.H. Schwarz, E. Witten,
``Superstring Theory'' Vol I, II

\end{thebibliography}
\end{document}